\documentclass[aps,prb,nobibnotes,reprint,superscriptaddress]{revtex4-1}
\usepackage{amsmath}
\usepackage{dcolumn}
\usepackage{graphicx}
\usepackage{float}
\usepackage{appendix}
\usepackage{multirow}
\usepackage{xcolor}
\usepackage{xr-hyper}
\usepackage{hyperref}

\begin{document}

\title{vdW-DF-ahcx: a range-separated van der Waals density functional hybrid}

\author{Vivekanand Shukla}
\email[Corresponding author: ]{vivekanand.shukla@chalmers.se}
\affiliation{%
Microtechnology and Nanoscience - MC2, Chalmers University of Technology, \\
SE-41296 Gothenburg, Sweden
}
\author{Yang Jiao}
\affiliation{%
Microtechnology and Nanoscience - MC2, Chalmers University of Technology, \\
SE-41296 Gothenburg, Sweden
}
\author{Carl Frostenson}
\affiliation{%
Microtechnology and Nanoscience - MC2, Chalmers University of Technology, \\
SE-41296 Gothenburg, Sweden
}

\author{Per Hyldgaard}
\email[Corresponding author: ]{hyldgaar@chalmers.se}
\affiliation{%
Microtechnology and Nanoscience - MC2, Chalmers University of Technology, \\
SE-41296 Gothenburg, Sweden
}

\date{\today}

\begin{abstract}
Hybrid density functionals replace a fraction of an underlying generalized-gradient approximation (GGA) exchange description with a Fock-exchange component. Range-separated hybrids (RSHs) also effectively screen the Fock-exchange component and thus open the door for characterizations of metals
and adsorption at metal surfaces. The 
RSHs are traditionally based on a robust 
GGA, such as PBE [PRL \textbf{77}, 3865 (1996)], for example, as implemented in 
the HSE design [JCP \textbf{118}, 8207 (2003)]. Here we define an analytical-hole (AH) [JCP \textbf{128}, 194105 (2008)] consistent-exchange (AHCX) RSH extension to the van der Waals density functional (vdW-DF) method [ROPP \textbf{78}, 066501 (2015)], launching vdW-DF-ahcx. We characterize the GGA-type exchange in the vdW-DF-cx version [PRB \textbf{89}, 035412 (2014)], isolate the short-ranged exchange component, and define the new vdW-DF hybrid. We find that the performance vdW-DF-ahcx compares favorably to (dispersion-corrected) HSE for descriptions of bulk (broad molecular) properties.
We also find that it provides accurate descriptions of noble-metal surface properties, including CO adsorption.
\end{abstract}


\maketitle

\section{Introduction}

The search for a better computational-theory understanding of small-molecule/organics-substrate interfaces is directly motivated by technological and environmental challenges. There is, for example, a need for interface insight to improve catalysts,\cite{BeGaLe15,catalysis1,ChJaGr19,AlAsNi20} batteries,\cite{lee2012li,lozano2017assessment,singh2020harnessing,shukla2017curious,shukla2019modelling}, gas adsorption \cite{vlaisavljevich2017performance,shukla2017toward}  and photocurrent generation in organic solar cells.\cite{RegGra91,YangCell13a,YangCell14,RanPRB16,Ingnas18,HoYaCu20} However, molecular adsorption on metal and semiconductor surfaces challenge density functional theory (DFT). There is only one defining approximation, namely in the choice of the exchange-correlation (XC) energy functional. However, that choice determines accuracy, transferability, and hence usefulness of the DFT calculations.\cite{BurkePerspective,beckeperspective} The problem for DFT practitioners lies in finding an XC choice that is optimal for descriptions at both sides of the interfaces.\cite{bearcoleluscthhy14,Interface_perspective,Tran19}

A van der Waals(vdW)-inclusive DFT approach is generally needed\cite{Interface_perspective} and we have to go beyond the traditional generalized gradient approximations (GGAs).\cite{lape80,lameprl1981,pewa86,pebuer96,PBEsol} Instead, 
one can use a ground-state energy functional with a dispersion correction,\cite{scoles,grimme1,becke05p154101,becke07p154108,silvestrelli08p53002,silvestrelli09p5224,ts09,Lilienfeld10p125010,grimme2,ts2,ts-mbd,grimme3,ambrosetti12p73101,umbd20,Jana20} a corresponding VV10-based extension,\cite{vv10,Bjorkman12p165109,Sabatini2013p041108,SCANvdW,solrVV10}
or move to the vdW density functional (vdW-DF) method.\cite{anlalu96,ryluladi00,rydberg03p126402,Dion04,Dion05,langreth05p599,thonhauser,lee10p081101,behy14,hybesc14,Berland_2015:van_waals,Thonhauser_2015:spin_signature,HyJiSh20,ChBeTh20}
The latter aims to track truly nonlocal correlation effects on the ground-state electron-density footing that DFT use to describe other types of interactions. There are many early 
examples of use for adsorption 
characterizations\cite{2001surfscience,rydberg03p606,chakarova-kack06p155402,johnston08p121404,toyoda09p2912,toyoda09p78,lan09p7165,langrethjpcm2009,toyoda10p134703,berland09p155431,moses09p104709,kong09p081407,wellendorff10p378,lee10p155461,lee11p193408,berland11p135001,kelkkanen11p113401,yanagisawa11p235412,rationalevdwdfBlugel12,le12p424210,Hamada2010p153412,lee12p424213,poloni_co2_2012,li12p121409,poloni_co2_2012,behy13,schroder13p871706,anti-aromatic,Bjork2014,AmbSil16} because the method grew out of a fruitful feedback loop involving many-body-perturbation theory (MBPT) and surface science.\cite{li56,dzlipi61,zarembakohn1976,zarembakohn1977,harrisandliebsch1982b,ma,harrisnordlander1984,li86,lavo87,anderssonetal1988,lavo90,andersson1993,dobdint96,dobdint96,anderssonpeha96,HolmApell,huanlula96,anhuryaplula97,anhuaplalu98,anry99,dowa99,huhyrolu01,schroder03p721,schroder03p880,schroder03p880,thonhauser,persson2008}
Meanwhile, use of a hybrid, i.e., mixing in a small fraction of Fock exchange, is also desirable. This is because hybrids are expected to improve the accounts of charge transfers.\cite{PBE0,HSE03,HSE06,KresseNJP,beckeperspective}

\begin{figure*}
\includegraphics[width=0.95\textwidth]{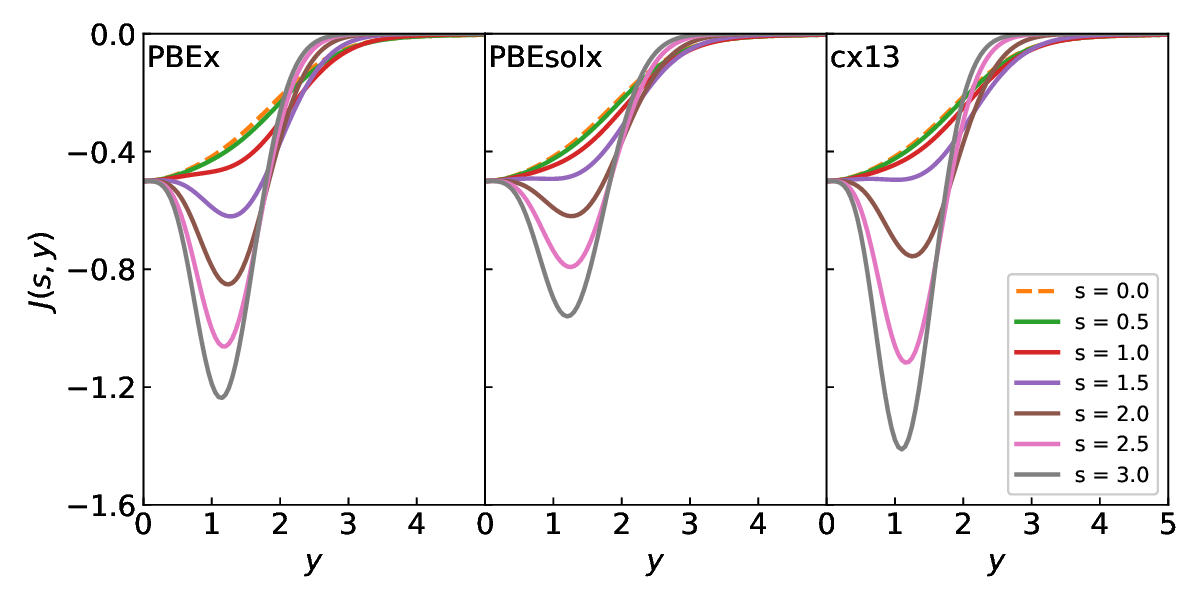}
\caption{\label{fig:newXset}
Shape of dimensionless exchange-hole 
$J(s,y)=n_{\rm x}(\mathbf{r},\mathbf{r'})/n(\mathbf{r})$ plotted against the (locally) scaled separation, $y=k_{\rm F}(\mathbf{r})|\mathbf{r}-\mathbf{r'}|$, between the electron at $\mathbf{r}$ and the `hole' (suppression of electron distribution) at $\mathbf{r'}$. The set of
curves reflects a range of assumed values of the local  scaled density gradient $s(\mathbf{r})$, from $s=0$
(in the homogeneous electron gas) to 3. 
The left, middle, and right panel characterize
the exchange in PBE (termed PBEx), in
PBEsol (termed PBEsolx), and in vdW-DF-cx (abbreviated CX and having the cx13, or LV-rPW86 exchange) in an analytical-hole (AH) formulation,\cite{HJS08} see text.}
\end{figure*}

There is broad experience that the use of hybrids and the inclusion of dispersion forces improves DFT accuracy, for example, in the description of molecules.\cite{gmtkn55,BurkePerspective,beckeperspective} Hybrids help us to correct for some self-interaction errors (SIEs).\cite{BurkeSIE,beckeperspective} The hybrids ameliorate a tendency of traditional, that is, density-explicit, functionals to overly delocalize orbitals. The VV10-LC\cite{vv10} and the vdW-DF0 class\cite{BerJia17,JiScHy18b} of hybrids are vdW-inclusive examples that rely on the ground-state electron density $n(\mathbf{r})$ to describe truly nonlocal correlations. The vdW-DF-cx0 design is based on the consistent-exchange vdW-DF-cx\cite{behy14,bearcoleluscthhy14,HyJiSh20} (below abbreviated CX) release of the vdW-DF method. The (unscreened) zero-parameter vdW-DF-cx0p (below abbreviated CX0p) hybrid\cite{JiScHy18b} uses a coupling-constant analysis of CX\cite{JiScHy18a,HyJiSh20} to set the Fock-exchange fraction $\alpha=0.2$ (within the vdW-DF-cx0 design). Compared with CX, the CX0p leads to significant improvements in the description of molecular reaction energies, particularly for small molecule properties. The CX0p is also accurate for the description of phase stability in the BaZrO$_3$ perovskites.\cite{PeGrRo20,jewahy20} On the other hand, the use of unscreened hybrids, like CX0p, is not motivated for metals and there is a clear argument that a non-unity dielectric constant in hybrids must be used even for 
molecules.\cite{BrGoVo17} Screening will certainly dampen the effects of long-range Fock exchange,\cite{HSE03,HSE06,HJS08,HSEsol,Koller_2013,OTRSHalga,OTRSHgap} for example, on the substrate side of the adsorption problem. 

In this paper we introduce a new range-separated hybrid (RSH), termed vdW-DF-ahcx, based on the consistent-exchange CX version, and therefore having the same correlation description as in the first general-geometry
vdW-DF version.\cite{Dion04,Dion05} Our new nonlocal-correlation RSH is named to emphasize that it constitutes an analytical-hole\cite{HJS08} (AH) consistent-exchange (AHCX) formulation of in the vdW-DF family. It will be abbreviated AHCX below. The design 
starts with an analysis of the CX exchange design, termed cx13 or LV-rPW86,\cite{behy14} that reflects the Lindhard screening logic and ensures current conservation.\cite{hybesc14,HyJiSh20} The design is similar to that of 
HSE\cite{HSE03,HSE06} and HSEsol\cite{HJS08,HSEsol} in that it focuses on isolating the short-range (SR) components, $E_{\rm x}^{\rm CX,SR}$ and $E_{\rm FX}^{\rm SR}$, of cx13 and of Fock exchange, respectively. As in HSE, we rely on an error-function separation,  
\begin{equation}
\frac{1}{r_{12}} = 
\frac{\hbox{erfc}(\gamma r_{12})}{r_{12}}    
+
\frac{\hbox{erf}(\gamma r_{12})}{r_{12}} \, ,
\label{eq:rscoulomb}
\end{equation} 
of the Coulomb matrix elements for electrons 
separated by $r_{12}=|\mathbf{r}_1-\mathbf{r}_2|$.
However, the HSE\cite{HSE03,HSE06} design was based on the
EP model of the PBE exchange hole.\cite{EP98} Here, we rely on the Henderson-Janesko-Scuseria (HJS) AH framework\cite{HJS08} for representing exchange in density functional approximations. This approach was also taken  in the definition of HSEsol, below discussed as HJS-PBEsol.\cite{HJS08,HSEsol} There are also HJS-based range-separated screened hybrids that include the long-range Fock exchange.\cite{ChaHea08,vv10,OTRSHalga,OTRSHgap,Jana19}
We discuss AH formulations of PBEx\cite{pebuer96} (in PBE), PBEsolx\cite{PBEsol} (in PBEsol), and cx13\cite{behy14} (in CX), to set the stage for defining AHCX.

The error-function separation ensures that AHCX screens the long-ranged component of Fock exchange. The AHCX is therefore set up to even describe metals and organic-molecular binding at metal (and high-dielectric-semiconductor) surfaces. The new RSH formulation aims to make the vdW-DF method available for bulk- and surface-application problems that require a more accurate description of charge transfer. The broad implementations of the adaptively compressed exchange (ACE) description\cite{linACE,PaoloElStruct1} 
in DFT code packages, such as \textsc{Quantum Espresso},\cite{QE,Giannozzi17} means that this strategy is also computationally feasible.

In fact, we find that AHCX is as fast as HSE for calculations of bulk and molecule properties. We document this for two bulk cases within, while we here provide an example comparison for medium-sized molecules, namely timing information extracted from our study of the C60ISO benchmark set on fullerene isomerizations, (Ref.\ \onlinecite{gmtkn55} includes a presentation of the set). These are just 10 out of the roughly 2300 molecular problems (investigated in multiple functionals) that are part of this AHCX launching work, but they give an impression. We study these 
C$_{60}$ systems in cubic unit cells of length 18.6 {\AA} using the ONCV-SG15\cite{ONCV,sg15} pseudopotentials (PPs)
at a 160 Ry wavefunction-energy cutoff. We average the total CPU-core-hour cost for completion across the ten geometries (although excluding a case where the Grimme-dispersion-corrected HSE-D3\cite{grimme3,gmtkn55} took an exceptional long convergence path) finding these timing results: PBE at 6 core hours, CX at 13 core hours, HSE-D3 at 568 core hours, and AHCX at 434 core hours. 

We note that the overall time consumption for hybrid studies of molecules is dominated by the 
\textsc{Quantum Espresso} ACE initialization. We also note that we have separately documented that the evaluation of the nonlocal correlation energy (used in AHCX and CX) scales well with cores and system size up to at least 10000 atoms, Ref.\ \onlinecite{libxcvdW}, and will not be a relevant bottle neck, at least not for a long time coming.

We motivate the AHCX design as a robust truly-nonlocal correlation RSH through the quality of the AH exchange description that we supply for cx13, that is, for the GGA-type exchange in CX. We argue that we can port the AH exchange description\cite{HJS08} from PBEx\cite{pebuer96} and PBEsolx\cite{PBEsol} to cx13, which is
constructed as a Pad{\'e} interpolation\cite{behy14} (LV-rPW86) of the Langreth-Vosko (LV) exchange\cite{lavo87,lavo90} and of the revised PW86 exchange.\cite{pewa86,mulela09} We note that related charge-conservation criteria are used in PBE/PBEsol and in the CX designs, and that they
all reflect the MBPT analysis of exchange in the weakly perturbed electron gas.\cite{mabr,rasolt,lape77,lape80,lavo87,lavo90,thonhauser} 

The rest of the paper is organized as follows. Section II summarizes the AH description of robust GGA exchange formulations, while Sec. III converts that insight into defining the AHCX. Section IV has computational 
details, a test of our AH-analysis approach, and a discussion of AHCX costs for bulk and other studies. Section V contains results and discussions, i.e., a documentation of performance for broad molecular properties, for bulk structure and cohesion, and for noble-metal surface properties, including CO adsorption. Sec.\ VI contains our summary 
and outlook. Finally, the appendix documents robustness of the adsorption-site-preference results with changes in the PP choice while the supplementary information (SI) material provides details of 
performance characterizations.

\section{GGA exchange and exchange holes}

The local Fermi wavevector $k_{\rm F}(\mathbf{r}) =
(2\pi^2n(\mathbf{r}))^{1/3}$ defines the (Slater) 
exchange energy density in the local density 
approximation (LDA), $\varepsilon^{\rm LDA}_{\rm x}(\mathbf{r})=-(3/4\pi) k_{\rm F}(\mathbf{r})$.
In a GGA, a local energy-per particle term also
defines the exchange part of the XC energy functional, 
i.e., the exchange energy functional
\begin{equation}
E_{\rm x}[n] = \int_\mathbf{r} n(\mathbf{r}) \, 
\varepsilon^{\rm GGA}_{\rm x}(n(\mathbf{r}); s(\mathbf{r})) \,  .
\label{eq:exchener}
\end{equation} 

The GGA energy-per particle description
$\varepsilon^{\rm GGA}_{\rm x}(n(\mathbf{r}); s(\mathbf{r}))$
depends on the electron density $n(\mathbf{r})$ and 
the scaled density gradient $s(\mathbf{r})= |\nabla n|/(2 k_{\rm F}(\mathbf{r}) n(\mathbf{r}))$. The ratio 
between the GGA and LDA energy-per-particle expressions defines the GGA exchange enhancement factor 
\begin{equation}
F_{\rm x}^{\rm GGA} (s(\mathbf{r})) = \varepsilon^{\rm GGA}_{\rm x}(\mathbf{r})/\varepsilon^{\rm LDA}_{\rm x}(\mathbf{r}) \, .
\end{equation}
As indicated, this enhancement factor can exclusively
depend on $s$, and the homogeneous electron gas (HEG) 
description is recouped by enforcing the limit value $F_{\rm x}^{\rm GGA} (s \to 0)=1$.

The exchange hole $n_{\rm x}(\mathbf{r}; \mathbf{r'})$ represents the tendency for 
an electron at position $\mathbf{r}$ to inhibit occupation of an electron of the same spin
at a neighboring point  $\mathbf{r'}$. It is part of the total XC hole $n_{\rm xc}(\mathbf{r};\mathbf{r'})$ which, in turn, is a full representation of all XC energy effects
\begin{equation}
    E_{\rm xc} = \frac{1}{2}\int_\mathbf{r}\int_\mathbf{r'} 
    \frac{n(\mathbf{r}) \, n_{\rm xc}(\mathbf{r};\mathbf{r'})}
    {|\mathbf{r}-\mathbf{r'}|} \, .
    \label{eq:ExcRel}
\end{equation}
This XC hole is defined via the screened 
electrodynamical response in the electron gas.\cite{ma,mabr,rasolt,gulu76,lape77,adawda,lape80,lavo87,lavo90,Dion04,thonhauser,HyJiSh20} It can 
be computed from the adiabatic connection formula\cite{gulu76,lape77} (ACF) and one 
can obtain a full specification the density-density fluctuations for the homogeneous 
electron gas (HEG) by MBPT or from quantum Monte Carlo calculations.\cite{lu67,Singwi68,Singwi69,helujpc1971,gulu76,Perdew_1992:accurate_simple}  Such 
calculations can, in turn, be used to establish a model for the HEG XC hole, $n_{\rm xc}^{\rm HEG}$, 
that takes a weighted or modified Gaussian form\cite{Singwi70,mahansbok} and defines LDA.\cite{gulu76,lape77,Perdew_1992:accurate_simple} 
Through the HEG, one can also extract and separately analyze the LDA exchange hole using a Gaussian model form.\cite{Perdew_1992:accurate_simple,EP98,HJS08}

The exchange hole for general electron-density distributions, 
\begin{equation}
\tilde{n}_x(\mathbf{r}; \mathbf{r'})  =  - 
\frac{n_1(\mathbf{r},\mathbf{r'}) \, n_1(\mathbf{r'}; \mathbf{r})}
{n(\mathbf{r})}\, ,
\label{eq:Fock}
\end{equation}
can, in principle,  be computed from inserting
single-particle orbitals in a one-particle density matrix 
\begin{equation}
n_1(\mathbf{r'},\mathbf{r}) = \sum_{i=1}^N \phi^{+}_i(\mathbf{r'})\phi_i(\mathbf{r}) \, .
\end{equation}
Using this $\tilde{n}_x$ form in Eq.\ (\ref{eq:ExcRel}) gives the Fock-exchange result
\begin{equation}
    E_{\rm FX} = \frac{1}{2}\int_\mathbf{r}\int_\mathbf{r'} 
    \frac{n(\mathbf{r}) \, \tilde{n}_{\rm x}(\mathbf{r};\mathbf{r'})}
    {|\mathbf{r}-\mathbf{r'}|} \, .
    \label{eq:ExxRel}
\end{equation}
However, a direct use of this Fock exchange in combination with the correlation parts of either LDA, GGAs or vdW-DFs is not desirable, for example, because it gives divergences in the description of extended 
metallic systems. In fact, a use of Fock exchange is also inappropriate for molecules and insulators, because electrons respond to and therefore screen the Coulomb field; The impact can be substantial also for molecules.\cite{BrGoVo17} Even a partial inclusion of  the
Fock-exchange description, Eq.\ (\ref{eq:ExxRel}), must 
(in general) be both compensated by correlation\cite{gulu76,lape77,lape80,lameprl1981,Levy85,Levy91,Gorling93,Burke97,Dion04,BurkePerspective,OTRSHgap,JiScHy18b} and described at an appropriate non-unity value of the dielectric constant.\cite{Koller_2013,OTRSHgap,OTRSHadsorp17,OTRSHalga,BrGoVo17}

Starting with the LDA description, one seeks instead an exchange-hole description $n_x$, and corresponding exchange energy functionals, Eq.\ (\ref{eq:ExcRel}), in which some XC cancellation is already built in.  The LDA, GGA, and vdW-DF descriptions for $n_{\rm x}(\mathbf{r}; \mathbf{r'})$,  and hence for $E_{\rm x}$, differ from the exchange description given by the Fock expression,  Eqs.\ (\ref{eq:Fock})-(\ref{eq:ExxRel}). In constraint-based
GGA, the assumption of modified Gaussian hole form is 
adapted (from the LDA start).\cite{pebuwa96,EP98,HJS08} 
This assumption also enters in a key role in the vdW-DF method by setting details 
of the underlying plasmon-response description.\cite{Dion04,lee10p081101,behy14,hybesc14,Berland_2015:van_waals,HyJiSh20} 

Since the GGA is given by the local value of the density $n$ and of the scaled gradient $s$, the GGA exchange hole must also be approximated in those terms and we introduce a dimensionless representation\cite{pewa86,mulela09,pebuwa96,EP98}
\begin{equation}
n_{\rm x}(\mathbf{r};\mathbf{r}') = n(\mathbf{r}) \, J_{\rm x}^{\rm GGA} (s(\mathbf{r}); y = 
k_{\rm F}(\mathbf{r})|\mathbf{r}-\mathbf{r}'|) \, .
\label{eq:GGAholeGeneral}
\end{equation}
Here the separation $|\mathbf{r}-\mathbf{r'}|$
is scaled by the local value of the Fermi 
wavevector $k_{\rm F}(\mathbf{r})$. The 
correspondingly scaled LDA exchange-hole form, $J_{\rm x}^{\rm HEG}(s)$, must arise as the proper $s\to 0$ limit 
of Eq.\ (\ref{eq:GGAholeGeneral}). 
For actual DFT calculations in the GGA, we need the resulting
exchange enhancement factor. We get it by inserting
the form of $J_{\rm x}^{\rm GGA}(n(\mathbf{r}),s(\mathbf{r}))$ 
into Eq.\ (\ref{eq:ExcRel}). The formal relation 
to the exchange energy $E_{\rm xc}$ is given 
by the enhancement factor
\begin{equation}
    F_{\rm x}(s) = - \frac{8}{9} \int_0^\infty y \, J_{\rm x}(s; y) \, dy\, .
    \label{eq:formalF}
\end{equation}

Perdew and co-workers furthermore defined a model for the GGA exchange hole, 
$J_{\rm x}^{\rm GGA}(n(\mathbf{r}),s(\mathbf{r}))$.
The weighted-Gaussian-hole 
form closely resembles that of $J_{\rm x} ^{\rm HEG}$,\cite{Perdew_1992:accurate_simple} 
but terms and exponents are now made 
functions of the scaled density gradient $s$.\cite{pewa86,pebuwa96,PBEsol,mulela09}
Parameters are set by charge conservation, constraints, and physics arguments. 
Such GGA models for the exchange hole have been defined for PW86/rPW86 exchange 
and PBE and PBEsol exchange descriptions (denoted PBEx and PBEsolx). The details 
in setting $J_{\rm x}^{\rm GGA}(n(\mathbf{r}),s(\mathbf{r}))$ produce different 
scaled-gradient enhancement factors, Eq.\ (\ref{eq:formalF}), and hence different exchange
energy functional descriptions, rPW86, PBEx, PBEsolx.

Connected with the PBE design, Ernzerhof and Perdew (EP) 
also designed an oscillation-free exchange-hole form,\cite{EP98} that simplified the definition of HSE as a PBE-based RSH.\cite{HSE03,HSE06} Signatures of Friedel
oscillations are present in, for example, the PBE exchange hole (in the $s\to 0$ limit),\cite{pebuwa96} 
but they are not considered physical for descriptions of molecules.  EP started by
a slight modification of the LDA description to extract a non-oscillatory LDA-exchange
hole form $J_{\rm EP,x}^{\rm LDA}(s)$ and then repeated the constraint-based GGA-exchange
design to craft the oscillation-free exchange hole form $J_{\rm EP,x}^{\rm GGA}$. It 
has been used to analyze both PBEx and PBEsolx.\cite{EP98,Constantin09} This EP representation 
for PBEx was used with range-separation, Eq.\ (\ref{eq:rscoulomb}), to establish HSE.\cite{HSE03,HSE06}

Characterizing the nature of the scaled exchange hole, Eq.\ (\ref{eq:GGAholeGeneral}) is equally relevant for the exchange components of vdW-DFs. This follows because the vdW-DF method presently relies on a GGA-type exchange, 
for example, cx13 (LV-rPW86) in the CX release.

\subsection{HJS model of PBE and PBEsol exchange holes}

Figure \ref{fig:newXset} shows exchange hole $n_{\rm x}$ descriptions for the semilocal PBE and PBEsol functionals, as well as for the truly nonlocal-correlation CX functional. The exchange part of the functionals are denoted PBEx, PBEsolx, and cx13. The hole shapes are here represented in the HJS or AH model framework for exchange \cite{HJS08},
using a form $J_{\rm x,HJS}^{\rm GGA}(n(\mathbf{r}),s(\mathbf{r}))$. 
However, the plots for PBEx and PBEsolx exchange can be directly compared with the EP-based analysis, $J_{\rm x,EP}^{\rm GGA}(n(\mathbf{r}),s(\mathbf{r}))$,\cite{EP98} shown in Fig. 2 of Ref.~\onlinecite{EP98} and Fig. 3 of Ref.~\onlinecite{Constantin09}; 
There are no discernible differences. There are also just small differences from the original PBEx and PBEsolx hole representations (apart from the removal of Friedel-oscillation signatures).

We work with the HJS framework for exchange holes because it gives several advantages. We call it an AH model of GGA exchange because it permits an explicit evaluation 
of Eq.\ (\ref{eq:formalF}) once $J_{\rm x,HJS}^{\rm GGA}(n(\mathbf{r}),s(\mathbf{r}))$ is inserted in
Eq.\ (\ref{eq:formalF}). The evaluation is stated
in Ref.\ \onlinecite{HJS08} and it also permits
a straightforward definition of functional derivatives.
This is a clear advantage when the AH analysis is used to also define and code (as summarized in the following section) a RSH hybrid HJS-PBE\cite{HJS08} that mirrors 
HSE. The new design simply 
involves 
changing from the $J_{\rm x,EP}^{\rm GGA}$ form to the $J_{\rm x,HJS}^{\rm GGA}$ form. 

A key AH advantage is ease of  generalization. It follows 
from having an analytical evaluation of Eq.\ (\ref{eq:formalF}). For example, HJS immediately established a plausible exchange-hole shape that reflects the PBEsol exchange enhancement factor.\cite{HJS08} In turn, this AH determination 
led to the definition of HJS-PBEsol,\cite{HJS08,HSEsol} that is, a RSH based on PBEsol. 
We shall use the AH-model for seeking generalization to 
the vdW-DF method in the following subsection.

\begin{figure}
\includegraphics[width=0.45\textwidth]{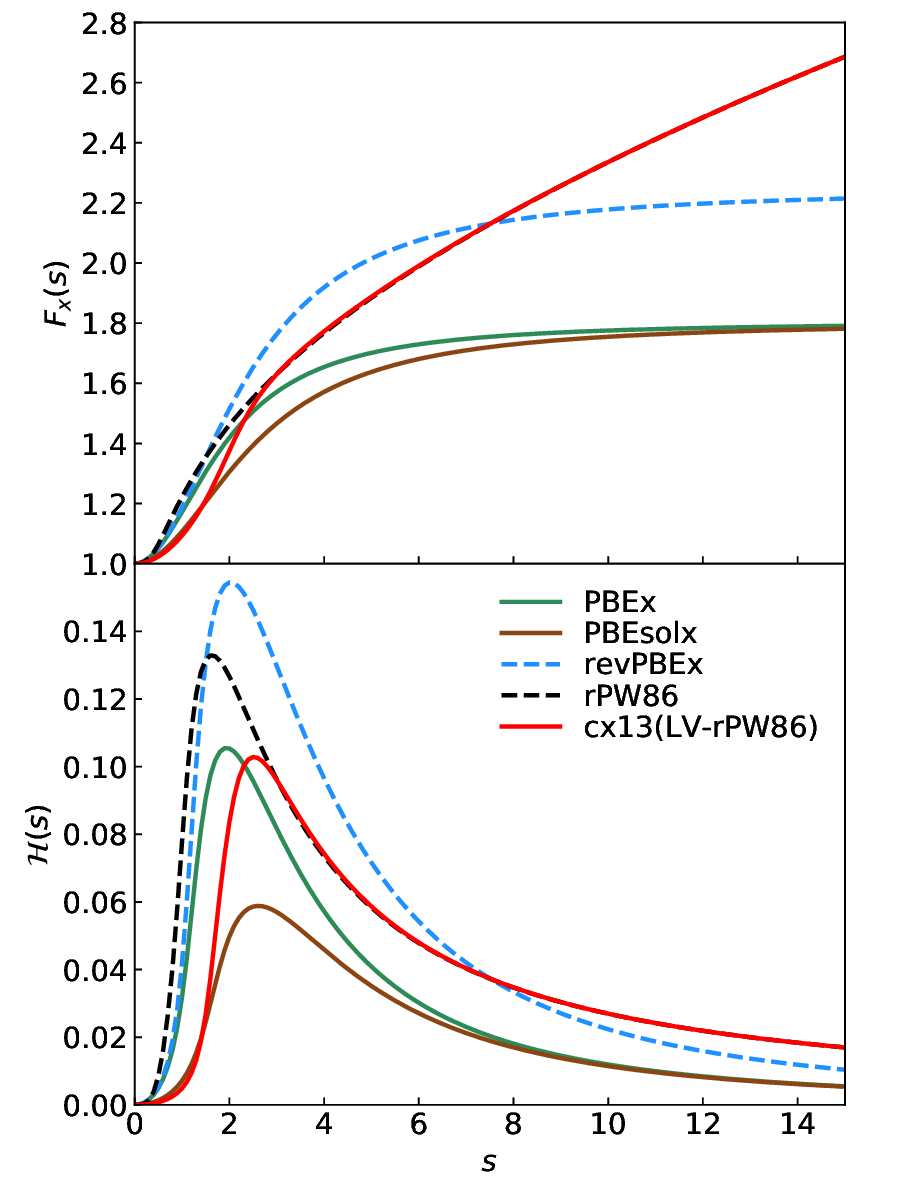}
\caption{\label{fig:FitHset}
\textit{Top panel:} The exchange gradient-enhancement functions $F_x(s)$ 
that reflect semilocal and truly nonlocal
functionals.  The last three exchange versions, 
denoted revPBEx, rPW86, and cx13 (that is, LV-rPW86)
are used in the vdW-DF1, vdW-DF2, and CX releases 
of the vdW-DF method.
\textit{Bottom panel:} Exponential factors $\mathcal{H}(s)$ that dominate in the description of corresponding exchange holes as obtained within the analytical HJS model description.}
\end{figure}

To begin, we summarize the HJS AH framework for characterizing the PBE and PBEsol exchange nature.
HJS revisits the mathematical framework for representing an LDA exchange hole
\begin{eqnarray}
    J_{\rm HJS,x}^{\rm LDA}(y) & = & - \frac{9}{4y^4} (1-e^{-\bar{\mathcal{A}}y^2})
    \nonumber \\
    & & + \left(\frac{9\bar{\mathcal{A}}}{4y^2} + \mathcal{B} + \mathcal{C}y^2 + \mathcal{E} y^4\right) 
    \, e^{-\mathcal{D}\,y^2} \, ,
    \label{eq:JHJS}
\end{eqnarray}
in a form that is free of signatures of Friedel oscillations.
The parameters $\bar{\mathcal{A}}, \mathcal{B}, \mathcal{C}, \mathcal{D}$ and $\mathcal{E}$ are
listed in the HJS reference \onlinecite{HJS08}. Like in the EP model, the parameters in Eq.\ (\ref{eq:JHJS}) are set by constraints on the exchange hole.

\begin{table}
\caption{Parameters of the rational function $\mathcal{H}(s)$ that set the Gaussian suppression in the exchange-hole shape in analytical-hole (AH) models\cite{HJS08} for PBEx,  PBEsolx,
and cx13 (i.e., the exchange in vdW-DF-cx).
The parameters are fitted against the numerically determined variation that results with a HJS-type analysis of the exchange functionals, see text and Ref.~\onlinecite{HJS08}.
\label{tab:paramH}
}
\begin{ruledtabular}
\begin{tabular*}{0.48\textwidth}{@{\extracolsep{\fill}}lrrr}
      &         PBE &      PBEsol &        cx13 \\
\hline
$a_2$ &      0.0154999  &      0.0045881  &      0.0024387  \\
$a_3$ &     -0.0361006  &     -0.0085784  &     -0.0041526  \\
$a_4$ &      0.0379567  &      0.0072956  &      0.0025826  \\
$a_5$ &     -0.0186715  &     -0.0032019  &      0.0000012  \\
$a_6$ &      0.0017426  &      0.0006049  &     -0.0007582  \\
$a_7$ &      0.0019076  &      0.0000216  &      0.0002764  \\
$b_1$ &     -2.7062566  &     -2.1449453  &     -2.2030319  \\
$b_2$ &      3.3316842  &      2.0901104  &      2.1759315  \\
$b_3$ &     -2.3871819  &     -1.1935421  &     -1.2997841  \\
$b_4$ &      1.1197810  &      0.4476392  &      0.5347267  \\
$b_5$ &     -0.3606638  &     -0.1172367  &     -0.1588798  \\
$b_6$ &      0.0841990  &      0.0231625  &      0.0367329  \\
$b_7$ &     -0.0114719  &     -0.0035278  &     -0.0077318  \\
$b_8$ &      0.0016928  &      0.0005399  &      0.0012667  \\
$b_9$ &      0.0015054  &      0.0000158  &      0.0000008  \\
\end{tabular*}
\end{ruledtabular}
\end{table}

The next step in the HJS exchange specification involves an extension to the case of gradient
corrections. In both the EP\cite{EP98} and the HJS\cite{HJS08} hole models, this gradient-extension form retains the Gaussian form. In the HJS form, the scaled exchange hole is written
\begin{eqnarray}
    J_{\rm HJS,x}^{\rm GGA}(s,y) & = & \mathcal{I}(s,y) \, e^{-s^2\mathcal{H}(s)\, y^2}
    \label{eq:ggaoverall} \\
    \mathcal{I}(s,y) & = &  \left(\frac{9\bar{\mathcal{A}}}{4y^2} + \mathcal{B} + \mathcal{C}\bar{\mathcal{F}}(s)y^2 + \mathcal{E} \bar{\mathcal{G}}(s)\, y^4\right) 
    \, e^{-\mathcal{D}\,y^2} 
    \nonumber \\
    & & - \frac{4}{9y^4} (1-e^{-\bar{\mathcal{A}} y^2}) \, .
    \label{eq:IHJS}
\end{eqnarray}
The $\bar{\mathcal{F}}(s)$ function
(affecting the $y^2$ term of the
LDA description) has the form
\begin{equation}
     \bar{\mathcal{F}}(s) = 1 -
     \frac{1}{27C} 
     \frac{s^2}{1+(s/2)^2} -
     \frac{1}{2C} s^2\mathcal{H}(s) \, .
\end{equation}
That is, it is given by the shape of $\mathcal{H}(s)$. Meanwhile, 
the final $\bar{\mathcal{G}}(s)$ modification (off of the LDA description) is set by an overall charge conservation criteria,\cite{gulu76,lape77,adawda,lape80,pewa86} i.e., a limit on the integral of the exchange hole. 

The important observation is that all components of the AH exchange description, Eq.\ (\ref{eq:IHJS}), are completely defined by the shape of $\mathcal{H}(s)$. The relation 
between $\bar{\mathcal{G}}(s)$
and $\mathcal{H}(s)$ can be explicitly stated in the HJS AH exchange model, see Eq.\ (40) of Ref.\ \onlinecite{HJS08}. Overall, the gradient-corrections provide, effectively, an extra Gaussian damping (defined by $\mathcal{H}(s)$), compared to the LDA description. The
suppression arises at large values of the scaled distance $y$. 
This damping is, however, offset by an enhancement of the exchange hole at and around $y=1$ (with the enhancement growing significantly with $s$). Nevertheless, all exchange hole components are completely set once we pick a plausible form for $\mathcal{H}(s)$.

In practice, the HJS AH modeling, for a given exchange of the GGA family, proceeds in 3 steps. First, we assume a rational form of the Gaussian damping function 
\begin{equation}
    \mathcal{H}(s) = \frac{\sum_{n=2}^7 a_n\, s^n}
    {1+\sum_{m=1}^9 b_m\, s^m} \, .
    \label{eq:Hform}
\end{equation}
and make a formal evaluation of Eq.\ (\ref{eq:formalF}):
\begin{eqnarray}
    F_{\rm x}(s) & = &
    \bar{\mathcal{A}} 
    - \frac{4}{9}
    \frac{\mathcal{B}}
    {\lambda}
    - \frac{4}{9}
    \frac{\mathcal{C}
    \bar{\mathcal{F}}(s)}
    {\lambda^2}
   - \frac{8}{9} 
    \frac{\mathcal{E}
    \bar{\mathcal{G}}(s)}
    {\lambda^3} 
    \nonumber \\
  & & + \zeta 
    \ln\left( \frac{\zeta}{\lambda} 
    \right)
    -\eta \ln\left( \frac{\eta}{\lambda}
    \right) \, ; 
\label{eq:FxsEval}
\end{eqnarray}
Here $\zeta = s^2 \mathcal{H}(s)$, 
$\eta = \bar{\mathcal{A}} + \zeta$,
and $\lambda = \mathcal{D} + \zeta$.
Second, since the GGA exchange is fully
specified by the enhancement factor $F_{\rm x}(s)$, we can invert Eq.\ (\ref{eq:FxsEval}) to get a numerical representation of $\mathcal{H}(s)$. Finally,
we fit that numerical determination to
the rational form, Eq.\ (\ref{eq:Hform}).

This HJS procedure formally works for any GGA exchange -- but it is an implicit assumption that we can still view the AH form $J$ as reflecting an implicit, constrained maximum-entropy principle.\cite{Perdew_1992:accurate_simple,pebuer96,pebuwa96,EP98} The Gaussian damping form,
$\mathcal{H}(s)$, must reproduce a given GGA-type  exchange enhancement factor $F_{\rm x}(s)$ while using as little structure as possible; There is no physics rationale for keeping such structure.\cite{pebuwa96,EP98} In a HJS representation, Eq.\ (\ref{eq:Hform}), we must strive to use as few significant polynomial coefficients as possible. We must also check that we have, indeed, avoided fluctuations that could hide non-physics input.

The left and middle panel of Fig.\ \ref{fig:newXset}
contrast the dependence on scaled density gradient
$s$ values of our AH models for PBEx and PBEsolx, as refitted here by us.  The AH model for PBEsol exchange has a softer variation with the scaled separation $y$ and with the scaled gradient $s$. In contrast, the AH for PBE exchange deepens more quickly. The GGA hole formulation is best motivated when the value of the scaled hole form $J$ remains larger than $-1$ (as it reflects a suppression of the electron occupation). The PBE and PBEsol exchange hole descriptions comply with that criteria for $s < 2.5$, i.e., in regions that are expected to dominate the description of bonding in materials and often also among molecules.\cite{pebuer96,pebuwa96,PBEsol,behy14,bearcoleluscthhy14,HyJiSh20,Ageo20}

The bottom panel of Fig.\ (\ref{fig:FitHset}) shows the shape of the exponential suppression or damping factor $\mathcal{H}(s)$ for PBEx and PBEsolx
(and for the cx13 exchange functional that is relevant for CX). For the PBEx analysis there is a close overlap with 
the exchange hole described in the original EP model, 
Ref.\ \onlinecite{EP98}, as also observed in the HJS 
paper, Ref.\ \onlinecite{HJS08}.  Similarly, 
the PBEsolx hole matches the EP-model-based 
representation asserted in Ref.\ \onlinecite{Constantin09}.

The AH hole characterizations  should be seen as trusted models of the PBEx and PBEsolx hole. The HJS AH analysis of PBEx and PBEsolx also  have nearly identical forms: Alone the scripted parameters denoted by an over bar in Eq.\ (\ref{eq:IHJS}) differs slightly in their formal expression. There are difference in the precise formulation of $\mathcal{H}(s)$. However, the resulting AH descriptions of  PBEx and PBEsolx can claim an important maximum entropy status because
of the close similarity with the EP descriptions.  

An important difference between the EP and the HJS exchange-hole descriptions lies in how these models are being used. The EP model\cite{EP98} was introduced to discuss the enhancement factor PBEx in terms of a  non-oscillatory exchange hole, and it led to HSE.\cite{HSE03,HSE06} The HJS model\cite{HJS08} allows for a simpler PBE discussion, but it is also being used for reverse engineering, i.e., being used to assert a plausible exchange-hole shape from the exchange enhancement factor. This track gave rise to both  HJS-PBEsol\cite{HJS08,HSEsol} and several recent long-ranged corrected hybrids.\cite{Jana19}

\subsection{Exchange-hole models for vdW-DFs}

This paper formulates the AHCX RSH (below) from an expectation that the HJS procedure (for reverse engineering an exchange hole form) also remains valid for both rPW86 and the LV exchange descriptions. This is plausible because cx13 is formed as a Pad{\'e} interpolation of the LV exchange\cite{lavo87,lavo90,thonhauser} (at small to medium values of the scaled density gradient $s$) and of rPW86\cite{pewa86,mulela09} (at large $s$ values). 

We first note that the PW86 exchange paper, Ref.\ \onlinecite{pewa86},
introduced a design strategy for setting the exchange hole within
the electron gas tradition. The procedure involves four steps: 1) Establish the symmetry limits on how the density gradient modifies the exchange hole off of the well-understood LDA representations; 2) establish the small-$s$ variation, for example, from MBPT results as in LV exchange;  3) impose an overall exchange-hole charge conservation criteria; and 4) extract the exchange enhancement factor from Eq.\ (\ref{eq:formalF}). Input beyond such physics analysis is always minimized in the electron-gas tradition that defines LDA, PW86/rPW86 exchange as well as both PBE, PBEsol and the vdW-DF method. 

We focus our wider AH analysis on the exchange choices that are used in the Chalmers-Rutgers vdW-DF releases. 
The exchange forms are the revPBEx (used in  vdW-DF1), 
the rPW86 (used in vdW-DF2), and cx13 (used in CX). 
One of these, rPW86,\cite{mulela09} is a refit of the 
PW86 that defined the strategy for making 
constraint-based 
GGA exchange design off of an model exchange hole.\cite{pewa86,pebuer96,PBEsol,HSE03,HJS08,HSEsol} We shall give a detailed motivation for trusting the 
HJS-type AH exchange characterizations for rPW86 
and cx13, below.

We have fitted the parameters $a_{2-7}$ and $b_{0-9}$ of Eq.\ (\ref{eq:Hform}) for 
the AH representations of the cx13, revPBEx, and rPW86. This extends our PBEx and PBEsolx discussion in the previous subsection. Additional
information is available in the SI material. 
Table \ref{tab:paramH} summarizes the most  
important such parameterization results: Our AH characterization for cx13 exchange and our AH refits of the HJS descriptions for PBEx and PBEsolx. 

The right panel of Fig.\ \ref{fig:newXset} provides a practical motivation for trusting our AH analysis of the cx13 (LV-rPW96)
exchange hole. The argument is given by noting similarities to the exchange-hole shapes of well-established exchange functionals. The cx13 shape begins (at low $s$) by reflecting a PBEsol nature and eventually it rolls over to the PBE-type (and rPW86-type) variation. Values greater than $s=2.5$ lead to the deeper hole minima, as in PBE. The cross over is not surprising since the PBEsol builds on a MBPT analysis that is close to the LV description (with small exchange enhancements up to medium $s$ values) while the rPW86 is known to have deep exchange holes. 

Figure \ref{fig:FitHset} summarizes 
the exchange enhancements (top panel)
and our AH analysis of corresponding
exchange holes (bottom panels). The top panel
confirms, in terms of exchange enhancements, that cx13 starts with a PBEsolx-(and PBEx-)like behavior at small $s$ values but transforms to the rPW86  behavior at large $s$ values. 

A more formal argument for trusting the AH analysis of cx13 exchange can be stated
as follows. We note that the CX leverages the formally exact electrodynamic-response framework of the vdW-DF method, as far as it is possible.\cite{HyJiSh20} The CX explicitly enforces current conservation  up to the cross over in the cx13 (or LV-rPW86) Pad{\'e} construction, namely at $s\approx 2.5$. As such, for the 
lower-$s$ LV end (up to 2-3), CX is an example of a consistent 
vdW-DF,\cite{HyJiSh20} systematically relying on a plasmon-based 
response description that adheres to all known constraints and sum rules.\cite{Dion04,Berland_2015:van_waals} At the LV end, cx13 furthermore leverages the Lindhard 
screening logic to  balance exchange and correlation.\cite{HyJiSh20} There is, in fact, 
a strong formal connection to the PBE constraint-based
design logic in that the CX emphasis on current conservation ensures an automatic compliance with the charge-conservation criteria.\cite{HyJiSh20} 

The cx13 or LV-rPW86 is a well-motivated electron-gas construction at this low-to-mid-$s$ LV
end. Here the cx-13 also satisfies the PBE-type local implementation\cite{pebuer96} of the Lieb-Oxford bound,\cite{LiebOxford81} 
systematically having exchange-enhancement 
factors $F_x(s) < 1.804$. For $s < 2.5$, the cx13 (and hence CX design) complies with the criteria on the exchange hole depth, $J > -1$.

At the high-$s$ end, the cx13 exchange design rolls over into the rPW86.\cite{mulela09} It is there naturally within the realm of  constraint-based exchange descriptions that started with PW86, Ref.\ \onlinecite{pewa86}. However, in this end, we must also discuss compliance with the actual, globally implemented, Lieb-Oxford bound.\cite{LiebOxford81} This is 
because the PW86 was designed before the importance of the bound 
was understood.\cite{pebuer96,pebuwa96,PBEsol}

For any given ground-state electron density $n$, we consider the ratio $R$ of the total exchange to the total LDA exchange (in the unit cell). Rigorous bounds are $R < 1.804$ (as is hard wired at the local level in PBE\cite{pebuer96}) and $R < 1.174$ for two-electron systems (as is hard wired 
at the  local level of SCAN\cite{SCAN,SCANvdW}). We note that
there are unphysical (spherical-shell) systems where the density is 
constructed so that the scaled 
density gradient is constant and where 
the actual Lieb-Oxford bound can only be 
satisfied if implemented also at the local 
level (as in PBE and SCAN).\cite{PeRuSuBu14}
For such unphysical systems, the rPW86 and cx13 
(and hence CX) will fail to comply with this 
exchange condition,\cite{LiebOxford81,PeRuSuBu14} if 
we furthermore assume that the unphysical system 
has a large scaled-density value, $s > 3$.

The important question remains, however, whether 
the cx13 (and hence CX) violates the actual 
Lieb-Oxford bound\cite{LiebOxford81} in real systems, 
that is, as encountered in actual DFT studies.\cite{Gharaee2017} 
The point is, that the Lieb-Oxford bound is formulated 
globally and must be checked on the unit-cell level in 
our periodic-system calculations.\cite{LiebOxford81} 

Answering this question is straightforward for any given
problem, as 
long as one saves the density after completing a \textsc{Quantum Espresso} calculation. Our now updated \textsc{ppACF} post-processing code (launched in  Refs.\ \onlinecite{JiScHy18a,JiScHy18b} and  committed to the \textsc{Quantum Espresso} package) gives a general coupling-constant analysis, an evaluation of kinetic-correlation energy components, as well as a per-system determination of actual exchange $R$ ratio (for nonhybrids). For the 10 fullerene calculations discussed in the introduction, we find that the cx13/CX value for the actual exchange ratio $R$ never exceeds 1.025. 
In practice, the CX exchange ratio remains far below even the stringent bound (1.174) that exists for two-electron systems.\cite{SCAN}

The bottom panel of Fig.\ \ref{fig:FitHset} provides further details of the cx13 design. It does so by comparing the Gaussian suppression factors $\mathcal{H}(s)$ that arise in our AH analysis of constrained exchange functionals. 
The shapes of the Gaussian suppression $\mathcal{H}(s)$ are similar in all cases. The $s$ position of the maximum for cx13 (LV-rPW86) sits essentially on the PBEsolx position, 
slightly above the maximum position 
for rPW86 and PBEx. Meanwhile, the cx13 maximum value 
aligns with that of the PBE description. As expected, the asymptotic cx13 behavior coincides with that of 
the rPW86 while for low $s$ values, the 
Gaussian suppression is almost identical
to that which characterizes PBEsol. This 
is also expected as the PBEsolx was explicitly designed to move the exchange enhancement closer to the input that also defines LV exchange. That is, the PBEsolx is closer to the MBPT analysis of exchange in the weakly perturbed electron gas.\cite{lavo87,lavo90,PBEsol,thonhauser,HyJiSh20}

Overall, Fig.\ \ref{fig:FitHset} confirms that there are no wild features in the
AH characterizations of exchange in the vdW-DF releases, including CX. Moreover, the bottom panel confirms our assumption, that our AH model description of cx13 can be trusted. This follows because we find that it
reflects a mixture of PBEsol-like, PBE-like,
and rPW86-like behaviors for the Gaussian suppression factor of the hole. In fact, like PBE, the cx13 aims to serve as a compromise of staying close to MBPT results at low $s$ (good for solids) and a more rapid enhancement rise (as in rPW86) for larger $s$ values (good for descriptions of molecular binding energies\cite{mulela09,kannemann}).

\section{Range-separated hybrids}

Hybrid functionals build on an underlying regular (density explicit) functional for XC energy. They simply replace some fraction $\alpha$
of the exchange description, $E_{\rm x}[n]$, of that functional with a corresponding component extracted from the  Fock-exchange term  
$E_{\rm FX}[n]$.
The latter is evaluated from Kohn-Sham orbitals
and calculations are carried to consistency in DFT.
We note that the regular vdW-DFs all have a GGA-type exchange by design and the vdW-DF hybrid design can therefore be captured in the same overall discussion. 

Simple, unscreened hybrids functional are described by the exchange component,\cite{PBE0,Perdew96,Gorling93,Ernzerhof97,Burke97}
\begin{equation}
    E_{\rm x}^{\rm hyb}[n] = \alpha E_{\rm FX}[n] + (1-\alpha)  E_{\rm x}^{\rm GGA}[n]\, ,
    \label{eq:hybrid}
\end{equation}
while the correlation term is kept unchanged. Here $E_{\rm FX}[n]$ is evaluated as 
the Fock interaction term (\ref{eq:ExxRel}). 
Examples of such  simple hybrids are PBE0 and the vdW-DF0 class.\cite{BerJia17,JiScHy18b} The CX0p results when we also use a coupling-constant scaling analysis\cite{Levy85,Levy91,Gorling93,Ernzerhof97,Burke97} to 
establish a plausible average value, $\alpha=0.2$, of the
Fock-exchange mixing in the vdW-DF-cx0 design.\cite{JiScHy18b}

For general RSH designs we split both the Fock exchange
and the functional exchange into short range (SR) and
long range (LR) parts:
\begin{eqnarray}
    E_{\rm FX} & = & E_{\rm FX}^{\rm SR}+E_{\rm FX}^{\rm LR} \, , \\
    E_{\rm x} & = & E_{\rm x}^{\rm SR}+E_{\rm x}^{\rm LR} \, . 
\end{eqnarray}
We simply insert the error-function separation, 
Eq.\ (\ref{eq:rscoulomb}), of the Coulomb matrix 
elements into Eq.\ (\ref{eq:Fock})
to extract, for example,  $E_{\rm FX}^{\rm SR}$. 

Given a trusted AH exchange representation
of the underlying functional `DF', we
can also extract the SR exchange component
\begin{eqnarray}
E_{\rm x}^{\rm SR,DF} [n] & = & \int_\mathbf{r} n(\mathbf{r}) \, 
\varepsilon^{\rm LDA}_{\rm x}(n(\mathbf{r}))
	F_{\rm x}^{\rm SR,DF}(k_{\rm F},s)  \, 
\label{eq:SRenergy}\\
	F_{\rm x}^{\rm SR,DF}(k_{\rm F},s) & = & - \frac{8}{9} \int_0^\infty y J_{\rm HJS}^{\rm DF}(s,y) \, 
\hbox{erfc}(\gamma y/k_{\rm F}) dy \, .
\label{eq:GenSRxEnhance}
\end{eqnarray}
Use of the EP model\cite{EP98} for a characterization of the PBE exchange hole, in combination with the range-separation Eq.\ (\ref{eq:rscoulomb}), led directly to the design of the first RSH, namely the HSE.\cite{HSE03,HSE06} 

More generally, for a trusted density functional `DF' (that have a GGA-type exchange functional part $E_{\rm x}^{\rm GGA}$,) we arrive at a HSE-type RSH or screened hybrid extension using
\begin{eqnarray}
    E_{\rm xc}^{\rm RSH,DF} & = & (1-\alpha) E_{\rm x}^{\rm SR,DF} + \alpha
    E_{\rm FX}^{\rm SR} \nonumber
    \\
    & & + E_{\rm x}^{\rm LR,DF} + E_{\rm c}^{\rm DF}
    \label{eq:ScreenedRSHdf}
\end{eqnarray}
where $E_{\rm c}^{\rm DF}$ is the correlation
part of `DF'. This correlation can be semi- or truly nonlocal in nature. The simple recast
\begin{equation}
    E_{\rm xc}^{\rm RSH,DF}[n] = E_{\rm xc}^{\rm DF}[n] + \alpha (E_{\rm FX}^{\rm SR} - E_{\rm x}^{\rm SR,DF}[n] ) \, ,
    \label{eq:hybridrecast}
\end{equation}
brings out similarities with the design of unscreened hybrids. 

For PBE and PBEsol there are already trusted AH descriptions in the HJS-PBE\cite{HJS08} and HJS-PBEsol\cite{HJS08,HSEsol} 
formulations (besides the original 
HSE\cite{HSE03,HSE06} obtained with the EP model-hole 
framework\cite{EP98}). To these we here add our own formulations of these analytical-hole screened hybrids, termed PBE-AH and PBEsol-AH. We do this 
to allow independent checks on our approach, as detailed
in the following section.

The main results of this paper are the definition and launch of the AHCX RSH. It is built from our AH exchange 
description for CX, again, as summarized in Table \ref{tab:paramH} and Figs.\ \ref{fig:newXset} and \ref{fig:FitHset}. The 
key observation is that we have trust in using the HJS type AH model of the CX exchange, so that we can rely on
Eq.\ (\ref{eq:GenSRxEnhance}) in establishing
an approximation for $E_{\rm x}^{\rm SR,CX}$.
We set the Fock exchange fraction at $\alpha=0.2$ as in the CX0p.\cite{JiScHy18b}
Still, the AHCX differs from CX0p by the presence of the inverse screening length $\gamma$ and thus a focus on correcting exclusively the short-range exchange description. We set
$\gamma=0.106$ (atomic units) as in the HSE06 formulation.\cite{HSE06}

The AHCX is a vdW-DF RSH that is constructed in the
electron gas tradition, being an all-from-ground-state 
density design.\cite{beckeperspective,hybesc14} That is, it
uses one and the same plasmon pole model\cite{behy14,HyJiSh20} 
for all but the Fock-exchange term. Even 
the Fock mixing can be seen as being set by the 
coupling-constant scaling of the consistent-exchange 
CX version, and thus given from within the 
construction.\cite{JiScHy18a,JiScHy18b} The AHCX is 
computationally more costly than meta-GGAs\cite{TPSS,revTPSS,SCAN,SCANvdW,Jana20} and 
DFT+U.\cite{AnZaAn91} These are other approaches 
that can also improve orbital descriptions 
and compensate for charge-transfer errors. 

The SCAN functional\cite{SCAN} is perhaps the alternative that is closest in nature
to CX and AHCX: they are all functionals of the electron-gas tradition and SCAN is constructed to retain some account of vdW interactions at intermediate distances\cite{SCAN} (thanks to 
input from a formal analysis of two-electron systems).

There are pros and cons of SCAN, CX, 
and AHCX use. SCAN and CX are certainly faster than AHCX.
However, SCAN must be supplemented by a separately-defined semi-empirical addition, in SCAN+rVV10,\cite{SCANvdW} to capture nonlocal-correlation effects across separations. 
In contrast, the AHCX is set up (as a single
XC functional design) to capture general interactions, for example, across the range of fragment separations at organics-metal interfaces.
One could extend the AHCX framework for use with optical or MBPT-specified tuning,\cite{Koller_2013,OTRSHgap,OTRSHalga,OTRSHadsorp17,BrGoVo17}
thus allowing for some motivated external parameters. However, the here-defined basic AHCX design is deliberately
kept free of adjustable parameters.

In practical terms, the SCAN, CX or AHCX choice
comes down to a discussion of the nature of the
material system as well as to attention to 
accuracy needs. Below we simply exemplify the 
AHCX potential in a set of demonstrator challenges (including noble-metal adsorption) and we include comparisons with CX and with literature
SCAN results\cite{patra2019rethinking,PaBaJi20} to 
illustrate differences in performance.

\section{Computational details and analytical-hole model validation}

We have coded the AH exchange-hole model and
AHCX (as well as HJS-PBE, PBE-AH, and PBEsol-AH) 
in an in-house version of \textsc{Quantum Espresso}.\cite{QE,Giannozzi17,PaoloElStruct1} 
It is a clear advantage of the HJS framework
that it allows an analytical evaluation
of the formal `SR' enhancement-factor expression
Eq.\ (\ref{eq:GenSRxEnhance}), for example, in
terms of finding and coding functional derivative
terms. The analytical evaluation is formally given in Eq.\ (43) of Ref.\ \onlinecite{HJS08}. For example, for our AHCX implementation we need simply to evaluate the terms for expressions and parameters specific to CX.

The implementation is fully parallel and can directly benefit from the computational acceleration that the ACE operator\cite{linACE,PaoloElStruct1} provides for the Fock-exchange component  $F_{\rm FX}^{\rm SR}$. This makes it possible to run efficient calculations of molecules, of bulk metals (requiring many $k$-points), and of surface slab systems on a standard high-performance-computer cluster.

For a demonstration of performance on bulk structure and cohesion, on broad molecular properties in the GMTKN55 suite,\cite{gmtkn55} and on CO adsorption, we use the electron-rich
optimized normconserving Vanderbilt\cite{ONCV} (ONCV) PPs, in the ONCV-SG15
release\cite{sg15}, at a 160 Ry wavefunction
energy cut off. 

Beyond the core documentation (bulk, molecule, and adsorption), we also include an AHCX demonstrator on workfunction and surface energy performance. These results involve computations of many slab geometries, all with surface relaxations,\cite{HyJiSh20} and at times with cumbersome electronic convergence. Hybrid studies 
of noble-metal surface properties are expensive
in the electron-rich ONCV-SG15 PPs. Meanwhile, use of norm-conserving PPs significantly helps 
stability of the ACE Fock-exchange evaluation that enters all types of hybrid calculations,
at least in the \textsc{Quantum Espresso} version where we placed our AHCX implementation. 
For the clean-noble-surface AHCX demonstrator work we therefore use the set of more electron-sparse AbInit normconserving PPs.\cite{abinit05} We argue for the validity of this approximation for noble metals and we track the likely impact by also providing workfunction and surface energy characterizations for PBE and CX in both the AbInit PPs and in the ONCV-SG15 PPs.

\begin{table}
	\caption{\label{tab:modelval} Comparison of atomization-energy results (in kcal/mol) obtained by the HSE, the HJS-PBE, and our PBE-AH description. The results is also compared against reference data. The comparisons among these closely related RSHs are made for the standard HSE choice of screening parameter $\gamma$ value and 
Fock-mixing value $\alpha$.
}
\begin{ruledtabular}
\begin{tabular}{lc|rrr|r}
                         &  
                         & HSE & HJS-PBE & PBE-AH
                         & Ref.$^a$\\
                         & $\alpha$  
                         & $0.25$   & $0.25$ & $0.25$ 
                         & -
                         \\
                         & $\gamma$ 
                         & $0.106$   & $0.106$ & $0.106$ 
                         & -
                         \\
\hline
propyne(C$_3$H$_4$)      &  &    699.9 &  699.27 &  699.23 & 705 \\
cyclobutane(C$_4$H$_8$)  &  &  1146.6 & 1145.51 & 1145.44 &   1149 \\
glyoxal(C$_2$H$_2$O$_2$) &  &    621.5 &  620.74 &  620.72 &     633 \\
SiH$_4$                  &  &    313.7 &  313.55 &  313.55 &      322 \\
S$_2$                    &  &   104.4 &  104.25 &  104.25 &    102 \\
SiO                      &  &    176.7 &  176.56 &  176.56 &  192 \\
\hline
$^a$Ref.\ \onlinecite{Curtiss97}.
\end{tabular}
\end{ruledtabular}
\end{table}

\subsection{Analytical-hole model validation} 

Table \ref{tab:modelval} reports
a simple sanity check that we provide to argue robustness of our maximum-entropy
description of the AH model and RSH constructions. We use the fact that Ref.\ \onlinecite{HJS08} reports
parameters for their AH 
characterization of PBE exchange,
a form that is here termed HJS-PBE.
We provide a modeling self-test using the ONCV-SG15 PPs at 160 Ry,
noting that the details of the $\mathcal{H}(s)$
specification must not change the
RSH results. 

Table \ref{tab:modelval} contrasts the results of such HSE-like descriptions for 6 atomization energies. For the HJS-PBE
and PBE-AH we chose values of
the Fock-exchange mixing $\alpha=0.25$
and for the screening $\gamma = 0.106$
(inverse Bohr's) that exist for default
HSE runs.\cite{HSE06} The agreement with reference data is good. The alignment of HSE and HJS-PBE (or AH-PBE) descriptions is very good.

Most importantly, Table \ref{tab:modelval} illustrates that our code to extract AH descriptions for PBE fully aligns results of PBE-AH
with those of HJS-PBE. We trust our
AH model description of PBE exchange, as defined by parameters in Table.\ \ref{tab:paramH}.

Interestingly, the details of our $\mathcal{H}$ specifications (Table \ref{tab:paramH}) differ
from those provided by HJS.\cite{HJS08} 
This is true even if there is no difference in the resulting RSH descriptions. This finding
suggests robustness in the
overall AH constructions of RSH,
not only for the HJS-PBE but also
for the here-defined vdW-DF-based
RSH, the AHCX.

\subsection{DFT calculations: molecules}

For a survey of performance on molecular properties
we have turned to the large parts of GMTKN55 that are 
immediately available for a planewave assessment
using the ONCV-SG15 PP set.\cite{ONCV,sg15}
The full GMTKN55 is organized into 6 groups of
benchmarks: 1) small molecule properties, 
2) large molecular properties, 3) 
barrier properties, 4) inter-molecular
noncovalent interactions (NCIs), 
5) intra-molecular NCIs, and 6) total NCIs. 
However, we exclude the G21EA and WATER27 
benchmark sets of GMTKN55.\cite{gmtkn55} In effect, we focus on a subset, the remaining easily-accessible `GMTKN53', in our discussion.

The G21EA benchmark is excluded because it contains negative charging of ions and
small radicals (like OH$^-$). Such small negatively charged systems genuinely challenge our planewave assessment of broad molecular properties exactly because we seek to maintain a high precision.
The fundamental problems arise by a combination of two factors. First, the self-interaction errors (SIEs) will, in 
these systems, push the highest-occupied molecular-orbital level up towards or above the vacuum floor, when it is done in a complete-basis set description.\cite{BurkeSIE} Second, our planewave description rapidly approached that complete-basis set limit due to our efforts to also carefully control spurious inter cell electrostatics interactions down towards the 0.01 kcal/mol limit.\cite{NeaLimit}
We conclude that a direct performance assessment on G21EA is meaningless in a planewave code and it is perhaps best left for a more advanced handling.\cite{BurkeSIE} 

We note in passing that we can approximately assert the G21EA performance by using the planewave equivalent of the so-called-moderate-basis-set approach,\cite{BurkeSIE} trapping the negatively charge ions in a small 
6-to-8-{\AA}-cubed boxes. We can thus obtain an approximate comparison that shows that the variation in G21EA performances has no discernible impact on the statistical performance measures that we track in GMTKN53. 

We furthermore removed WATER27 for a 
related SIE-induced\cite{BurkeSIE} problem - the
\textit{path} to electronic-structure convergence for the OH$^-$ is exceedingly cumbersome once we seek to \textit{fully} converge the WATER27 set with respect to unit-cell size. Of course, hybrids helps in the WATER27 (and in the G21EA) study -- but then we can offer no fully converged comparisons. Again, using as separate (12 {\AA}) small-box handling of the OH$^-$ system makes it possible to complete an approximate WATER27 benchmarking; As will be detailed elsewhere, this assessment identifies vdW-DF2, CX, CX0p and AHCX as top performers.

Fortunately, having 53 (out of 55 sets of Ref.\ \onlinecite{gmtkn55}) still makes for ample statistics. 
Accordingly, we simply work with the easily-accessible GMTKN53 
and compare with literature GMTKN55 results.\cite{gmtkn55} This
is fair because, if anything, we have negatively offset
our AHCX performance reporting (by omitting G21EA and WATER27).

We supplement the GMTKN53 characterization both 
by tracking performance differences (that arise for some individual benchmarks) and by analyzing the groups of benchmarks that mainly reflect NCIs. We essentially use the same computation setup as the pilot study included in a recent focused review of consistent vdW-DFs, Ref.\ \onlinecite{HyJiSh20}. However, we have now 
moved off of the electron-sparse AbInit PPs and onto the electron-rich ONCV-SG15;\cite{ONCV,sg15} Moreover,
we now make systematic use of large cubic cells 
and a Makov-Payne electrostatics decoupling;\cite{Makov} In the pilot assessment,\cite{HyJiSh20} the electrostatic decoupling was only done for charged systems.

We find that this upgrade of our benchmarking strategy accounts for some of the deviations that we previously incorrectly ascribed to limitations of the CX\cite{behy14} and vdW-DF2-b86r\cite{hamada14} functionals.\cite{HyJiSh20} The past limitations seem to, relatively speaking, especially impact our assessment of Group 4  (intermolecular NCI) performance, as represented in Refs.\  \onlinecite{HyJiSh20,Jana20}. This paper provides a more fair performance characterization of 
the performance of CX and vdW-DF2-b86r as well as 
of AHCX relative to other vdW-inclusive DFT 
versions, for example, dispersion-corrected metaGGA.

\subsection{DFT calculations: bulk and surfaces} 

For bulk structure and cohesive energies, we used the Monkhorst-Pack scheme with an $8\times 8 \times 8$ $k$-point grid for the Brillouin-zone integration. 
Our demonstration survey includes 4 semiconductors, 
3 ionic insulators, 1 simple metal, and 5 transition 
metals including Cu, Ag, and Au. Here we also used the electron-rich ONCV-SG15\cite{ONCV,sg15} 
PPs set at 160 Ry wavefunction energy cut-off. 
We tracked the total-energy variation with
lattice constant and used a fourth-order polynomial
fit\cite{ZiSc03} to identify optimal (DFT) structure,
the cohesive energy, and bulk modulus.

Surface energies for Cu(111), Ag(111), and Au(111) were estimated from the energies calculated from 4 to 12 layers of the noble metals in unit-cell configurations with 15 {\AA} vacuum. Here we sought to offset some of the high computational cost by instead using the AbInit normconserving PPs\cite{abinit05}
for extra demonstrations of AHCX usefulness. In the surface 
studies, we first determine the in-surface unit cell from the calculated bulk lattice constant. Next we allow the outermost
atoms to relax in a slab-geometry description. The exception is for AHCX 
characterizations, where we instead relied on 
the results of the CX structure characterizations.

To document AHCX accuracy for CO adsorption on noble-metal surfaces, we used again the electron-rich ONCV-SG15 PPs at 160 Ry. We consider the FCC and TOP sites in a 2x2 surface-unit-cell description and a six-layer slab geometry. We also track the adsorption-induced relaxations on the top three layers and compute the CO adsorption energy as follows,
\begin{equation}
E_{\rm ads}= E_{\rm CO/metal(111)}-E_{\rm{metal}(111)}-E_{\rm CO} \, .
\end{equation}
Here $E_{\rm CO/metal(111)}$ denotes the fully relaxed energy of the adsorbate system, while $E_{\rm metal(111)}$ and $E_{\rm CO}$ denote the total energy of the fully-relaxed (isolated) surface and molecule system, respectively.

\subsection{Computational costs}

\begin{figure}[t]
\centering
\includegraphics[width=1.0\linewidth]{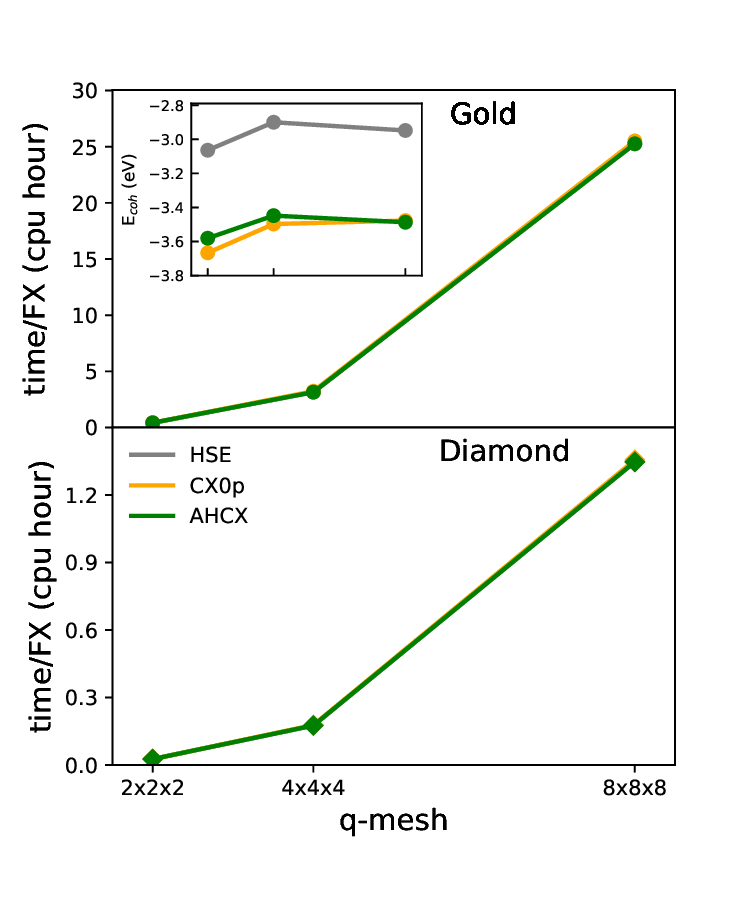}
\caption{\label{fig:Timescale} Scaling of computational cost for hybrid calculations of gold and diamond using the primitive cell. We compare central-processing-unit(cpu)-hour cost per Fock-exchange evaluation step for calculations of the bulk cohesive energies as computed using a $8\times 8 \times 8$ $k$ point mesh and 160 Ry wavefunction energy cut off for the ONCV-SG15 PPs. We track the dependence on using a so-called `down-sampled' $q$ mesh (setting $k$-point differences) for evaluating the Fock-exchange contribution in the hybrid study. The inset tracks 
how the cohesive energies change with the choice of the $q$ mesh.}
\end{figure}

Figure \ref{fig:Timescale} and Table S.II of the SI material detail the computation time required for completing one evaluation of the ACE Fock-exchange operator in hybrid calculations on a single
20-core node for gold and diamond, top and 
bottom panels, respectively. We include a 
comparison of HSE,\cite{HSE06} CX0p,\cite{JiScHy18b} 
and AHCX for gold and diamond, using a 8x8x8 $k$ point grid and the electron-rich ONCV-SG15 PPs. We track the variation with the so-called $q$ mesh (of $k$ point differences used in the Fock exchange evaluation).

There is no additional cost of doing AHCX over 
HSE (as in the molecular cases discussed in the introduction) and the AHCX cost is slightly
lower than the cost of the corresponding 
unscreened hybrid CX0p. This is in part a statement of the computational efficiency of the new nonlocal-correlation RSH functional. It is also a statement that we have provided a robust implementation of the analytical hole description and that we have succeeded in leveraging the benefits of the \textsc{Quantum Espresso} ACE implementation.\cite{PaoloElStruct1}

An overview of scaling of AHCX (or HSE) hybrid calculations can be deduced from Ref.\ \onlinecite{PaoloElStruct1}. The cost
per Fock exchange evaluation scales with the FFT scope (size of unit cell) and with the number of bands (that is, number of electrons per unit cell) squared.\cite{PaoloElStruct1} Also, for
hybrid studies of bulk and surfaces one needs to consider multiple $k$ points and a $q$ mesh of $k$-point differences, causing an additional scaling factor. When there is symmetry (as in simple bulk) one can cut down this factor significantly by limiting the set of $k$ and $q$ points that enters in an actual Fock-exchange evaluation. Importantly, however, there are communication costs when jobs are spread across 
multiple nodes.\cite{PaoloElStruct1} For example, we can push more hybrid calculations with electron-sparse PPs because they require fewer bands; By limiting bands we both directly accelerate the computation speed and we reduce the memory requirements, making our jobs fit on more types of computer resources.

The AHCX is more costly than non-hybrids, including metaGGAs, for example, as illustrated in the introduction. Also, as mentioned above, hybrid calculations are costly for surfaces where relaxations enviably produce low-symmetry geometries. However, the AHCX computational costs can be handled by choosing an appropriate parallelization strategy. We try to limit the number of nodes as we also try to accommodate the memory requirements.

For a specific illustration of AHCX costs we make the following observations from our studies. We report summary of nearly 5000 AHCX and HSE-type hybrid studies of molecular properties of the GMTKN55 (having up to 37-{\AA}-cubed unit cells to carefully control electrostatic couplings); Those ONCV-SG15  jobs took us a total of 200000 core hours (5-6 times 
the cost of using the corresponding CX and PBE 
regular functionals). The cost for an individual bulk structure study in AHCX/HSE is (due to symmetry) relatively cheap (see Fig.\ \ref{fig:Timescale});  Here the 700+ computations of energy-versus structure variations that enters our structure optimization (of 11 simple cases and of 5 transition metals) took just over 100000 hours at the production stage. Finally, the cost for hybrid studies of noble-metal surface energies and adsorption are on a different scale. We were able to obtain the 6 frozen-geometry AHCX CO-noble-metal adsorption energies in the electron-rich ONCV-SG15 PP setup at the costs of about 100000 core-hours each (on selected computer resources). For the additional noble-metal workfunction and surface energy demonstrators (involving many slabs) we cut the cost to about 1.5 million core hours by instead relying on the electron-sparse AbInit PPs for AHCX calculations.

\section{Results and discussion}

\subsection{Broad molecular properties}

\begin{figure}[t]
\centering
\includegraphics[width=1.0\linewidth]{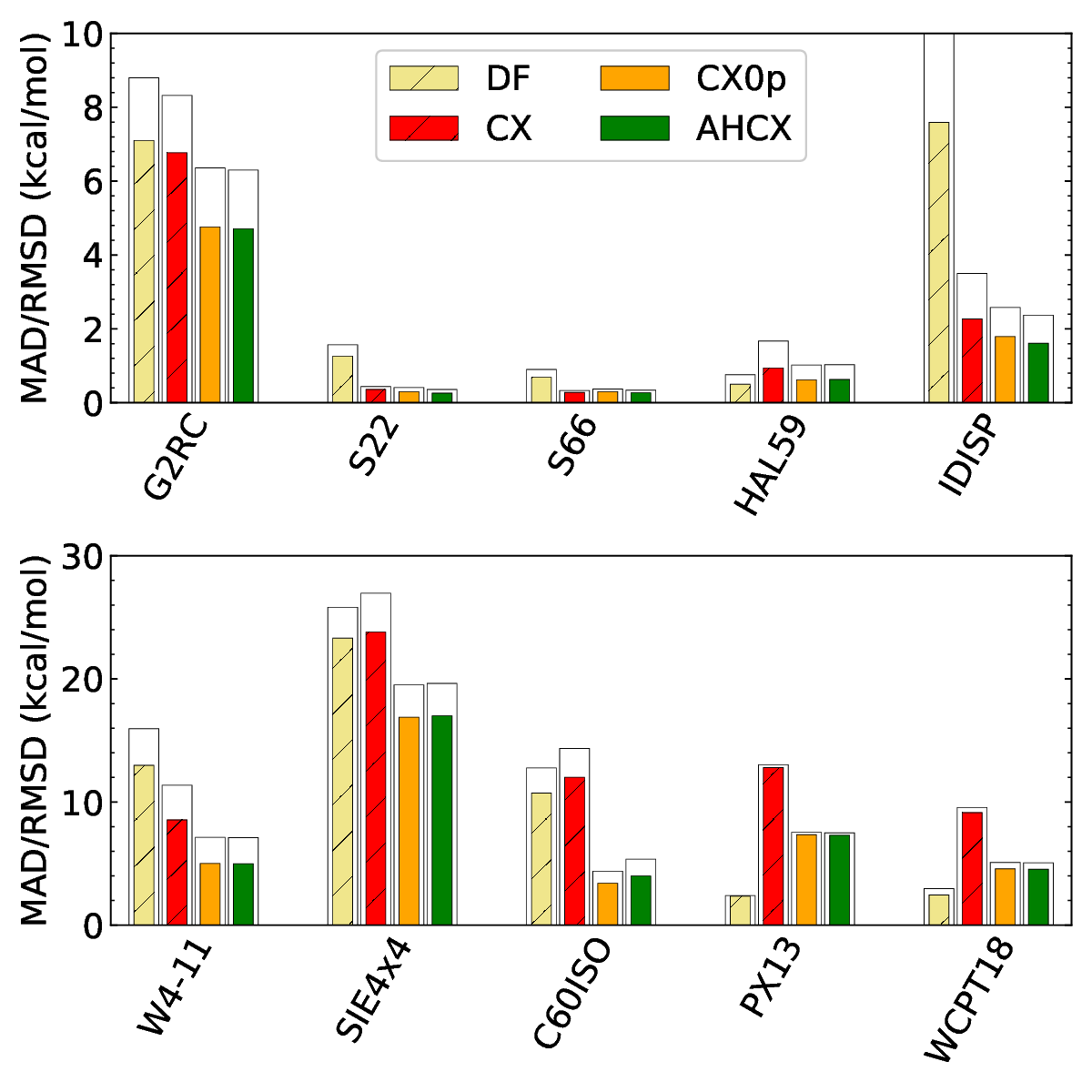}
\caption{Comparison of vdW-DF (sand), of CX (red), of CX0p (orange), and AHCX (green) performance in various individual molecular benchmark sets of the GMTKKN55 suite. The histogram bars represent mean absolute deviation (MAD) and root-mean square deviations (RMSD) relative to quantum chemistry reference data (at reference geometries) as listed in the GMTKN55 suite for broad benchmarking on molecular
properties. The upper panel contrasts the performance 
for small molecule performance (G2RC),  and for the (intra- plus inter-molecular) noncovalent-interaction benchmark sets (S22, S66, HAL59, and IDISP).  The lower panel provides a performance comparison for benchmark sets reflecting molecular atomization energies (W4-11), challenging SIE problems (SIE4x4), C60 isomers (C60ISO) as well as proton-transfer barriers (PX13 and WCPT18).}
\label{fig:MolValidation}
\end{figure}

Figure \ref{fig:MolValidation} shows a 
performance comparison that we provide for 10 molecular benchmarks of the full GMTKN55 suite. The figure characterizes the performance in terms of mean absolute deviation (MAD) values. Use of hybrid AHCX provides significant improvement over CX in the case of self-interaction problems (in the SIE4x4 benchmark set), for atomization energies (in the W4-11) set, and for several barrier-type problems, including those reflected in the PX13 and WCPT18 benchmark sets. The improvements are perhaps largest in the case of C60 isomerizations, for large deformations having stretched bonds. 

\begin{figure}[t]
\centering
\includegraphics[width=1.0\linewidth]{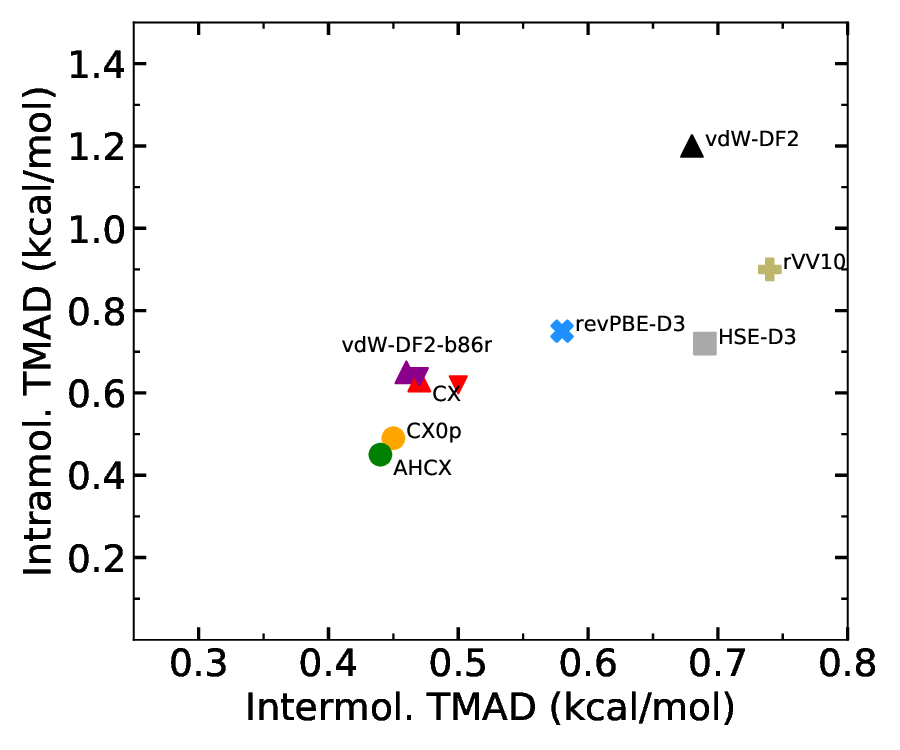}
\caption{\label{fig:NOCscatter}
Progress in performance from
regular vdW-DFs (upward triangles) to hybrid vdW-DFs (circles) on benchmark sets reflecting noncovalent interactions (NCI), but excluding the WATER27 set. The SI material documents that vdW-DF1 can be found just outside the range. Our characterization can be directly compared to the broader survey reported in Fig. 3 of Ref.\ \onlinecite{Jana20}. As in that study, the functional
performance is represented in terms of `total mean-absolute-deviation' (TMAD) values (that average
MAD values over benchmark sets, see text)
obtained for intermolecular NCI sets (x-axis)
and for intramolecular NCI sets (y-axis). The present characterizations for CX and vdW-DF2-b86r are more accurate than the assessment (shown by downward triangles) previously obtained in a pilot characterization.\cite{HyJiSh20}  
For reference we also include  characterizations of dispersion corrected revPBE-D3 and HSE-D3, as well as of rVV10.}
\end{figure}

Figure \ref{fig:NOCscatter} summarizes our survey of 
performance across the entire NCI groups of the GMTKN55,
although excluding the WATER27 benchmark set for reasons explained above. The figure characterizes the performance in terms of a total MAD, or TMAD, value for intermolecular NCIs (abscissa position) and of a TMAD value for intramolecular NCIs (ordinate position). These effective MAD values result as we first compute the MAD values for each NCI benchmark of the GMTKN55 and then average over these MADs over the number of  
intermolecular and intramolecular NCI benchmarks investigated in GMTKN55 Group 4 and Group 5, as in Ref.\ \onlinecite{Jana20}. The SI material provides details and lists the quantitative data that we provide for vdW-DF hybrids (AHCX and CX0p, circles) and for regular vdW-DFs (vdW-DF2, vdW-DF2-b86r and CX, upwards triangles), as well as for rVV10 and dispersion corrected revPBE-D3 and HSE-D3. The SI material
also includes a performance characterization
of vdW-DF1\cite{Dion04,Dion05,thonhauser}
and vdW-DF-ob86.\cite{vdwsolids}

The NCI assessments in Fig.\ \ref{fig:NOCscatter} can be compared to Fig.\ 3 of Ref.\ \onlinecite{Jana20},
a study that also summarized NCI results
obtained in Refs.\ \onlinecite{gmtkn55,NajGoe18}.
The assessments differ in code nature
(planewave versus orbital-based DFT)
but we can use Fig.\ \ref{fig:NOCscatter} for a broader discussion of performance. That is, our comparison for NCI performance also indicates how the new AHCX RSH vdW-DF hybrid fares in relation to the dispersion-corrected metaGGAs and dispersion-corrected hybrids discussed in Ref.\ \onlinecite{Jana20}. We find that
the AHCX performance is better than those of the dispersion-corrected metaGGAs and of $\omega$B97X-D3
(as reported in Ref.\ \onlinecite{gmtkn55,NajGoe18}).
However, our AHCX does not fully
match the performance of DSD-BLYP-D3,\cite{gmtkn55}
when it comes to the 
TMAD value for intermolecular NCI.

For completeness, Fig.\  \ref{fig:NOCscatter} also shows the approximate assessments (downwards triangles) of CX and vdW-DF2-b86r performance that we obtained in a
pilot benchmarking.\cite{HyJiSh20} The differences in CX and vdW-DF2-b86r positions of upwards and downward filled triangles reflect the improvement that we have here made in benchmarking strategy relative to Ref.\ \onlinecite{HyJiSh20}. The 
systematic use of decoupling of spurious dipolar
inter-cell attractions is found to have a clear effect on performance for the group of intermolecular NCI benchmark sets.

We find that the use of vdW-DF hybrids leads to significant improvements in the description of 
problems characterized by strong NCIs. The performance of AHCX and CX0p are again nearly identical but significantly better than dispersion-corrected 
HSE-D3 and of the regular vdW-DFs. 
This is not surprising because the hybrids are expected to have a more correct description of orbitals and reshaping orbitals (and hence the
density variation)  will directly affect the strength of the vdW attraction.\cite{lo30,lo37,kleis07p100201,ts09,asymptoticseries,fullerenewisdom,C60crys}

\begin{figure}[t]
\centering
\includegraphics[width=1.0\linewidth]{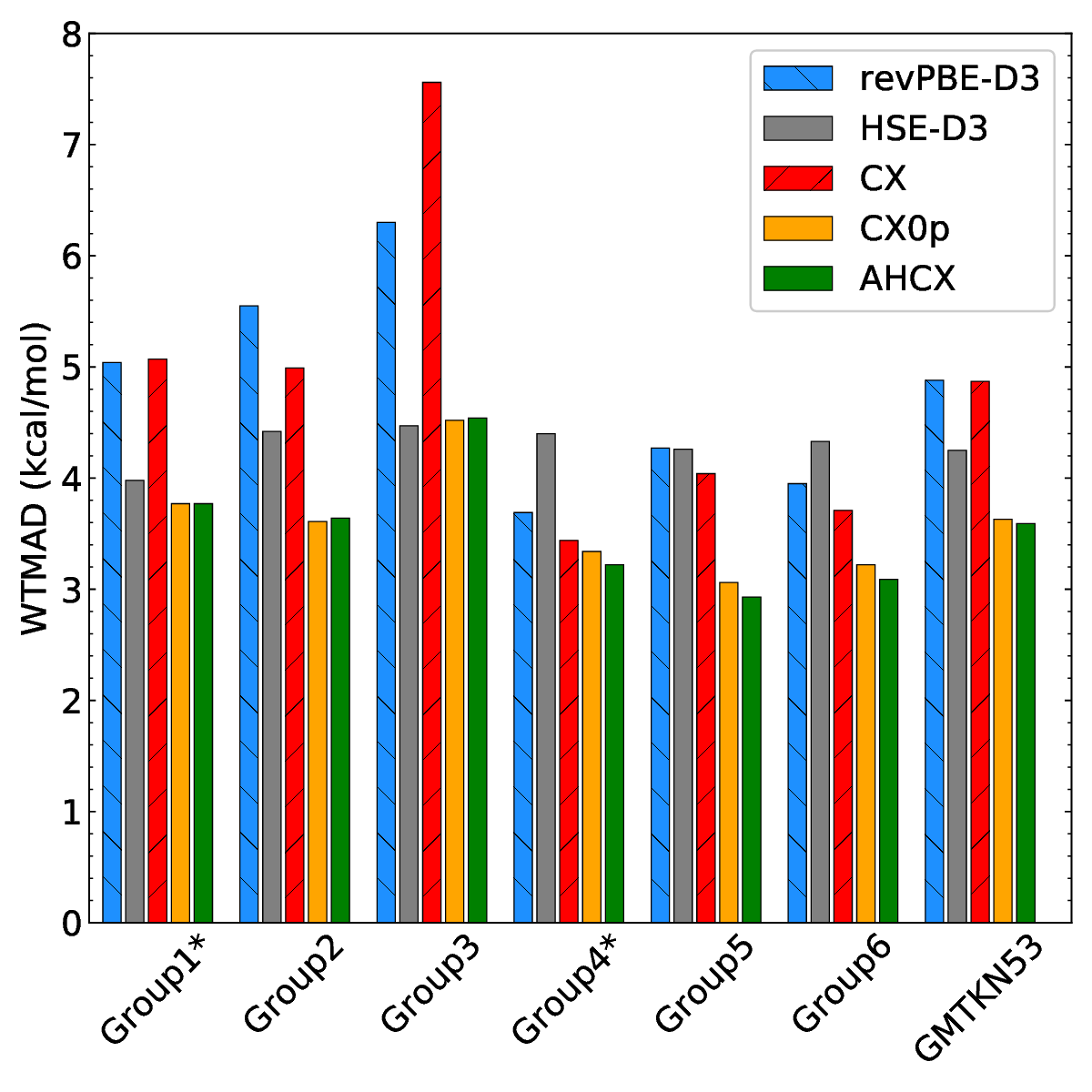}
\caption{\label{fig:GMTKN53comp} Comparison of molecular performance of revPBE-D3 (a strong-performing dispersion-corrected GGA), of HSE-D3 (a dispersion-corrected PBE-based RSH) and of the family of consistent vdW-DFs \cite{HyJiSh20}. This family now includes CX (a regular vdW-DF release), CX0p (a corresponding unscreened hybrid), and AHCX (a CX-based RSH). The comparison is given for a 53-benchmark
subset of the full GMTKN55 suite,\cite{gmtkn55} namely those
that are directly available to planewave benchmarking using the electron-rich ONCV-SG15 PPs. As indicated by the `*' superscripts, we have here omitted the G21EA benchmark set from the GMTKN55 Group 1 (small molecular properties) and the WATER27 benchmark set from Group 4 (intermolecular-noncovalent interactions) because they contain small negatively charged ions and radicals (see text); We include all benchmark sets from Group 2 (large molecular properties), Group 3 (barrier properties), and Group 5
(intramolecular properties), and thus essentially all sets in Group 6 (total noncovalent interactions). The performance is measured by the weighted total mean absolute deviation WTMAD1 measure introduced in Ref.\ \onlinecite{gmtkn55}.}
\end{figure}

For a summary of overall progress that the hybrid vdW-DFs may bring for general molecular properties, we rely on
the `easily available GMTKN53', defined above. This is the  subset  of the GMTKN55 suite that omits the WATER27
and G21EA benchmarks sets, for good reasons.\cite{BurkeSIE}
Fortunately, with GMTKN53 we remove just one benchmark set
in two groups and can still provide a balanced account.
It is still meaningful to discuss functional performance in terms of per-group results as in the GMTKN55 study.\cite{gmtkn55} 

For comparison of performance we therefore use the reference geometry and reference data of Ref.\ \onlinecite{gmtkn55} to determine the deviation
on each of the over 2000 molecular processes that are included. We evaluate the weighted total mean absolute deviation measures 
(WTMAD1 and WTMAD2) that Grimme and co-workers introduced
to allow a comparison also among the GMTKA55 groups.\cite{gmtkn55} The adaptation of these
measures to the present GMTKN53 survey simply
involves adjusting the counting of 
participating benchmark sets for GMTKN55
groups 1 and 4,\cite{gmtkn55} see SI material
for further details.

Figure \ref{fig:GMTKN53comp} shows our summary functional comparison based on this `easily accessible GMTKN53', using WTMAD1 measures. The SI material lists the data for both WTMAD1 and WTMAD2 comparisons. We find that the AHCX performance is significantly better
than dispersion-corrected HSE-D3 for this assessment of broad molecular properties. The AHCX also performs 
systematically better than CX, which in turn 
performs at the level of dispersion-corrected 
revPBE-D3. The latter is reported to be one of the
best performing dispersion corrected GGAs.\cite{gmtkn55}

Importantly, the AHCX also performs at the same level and perhaps slightly better than the unscreened CX0p hybrid. Our primary motivation for defining AHCX is to have a vdW-DF RSH that can also describe adsorption at metallic surfaces, which is not something that the unscreened CX0p is set up to do.  
The comparison in Fig.\ \ref{fig:GMTKN53comp} shows
that the AHCX is also good enough on the molecular side
of interfaces. That is, it remains a candidate for also
serving us for the interface problems, at least so far.

\subsection{Bulk structure and cohesion}

\begin{figure}[t]
   \centering
   \includegraphics[width=1\linewidth]{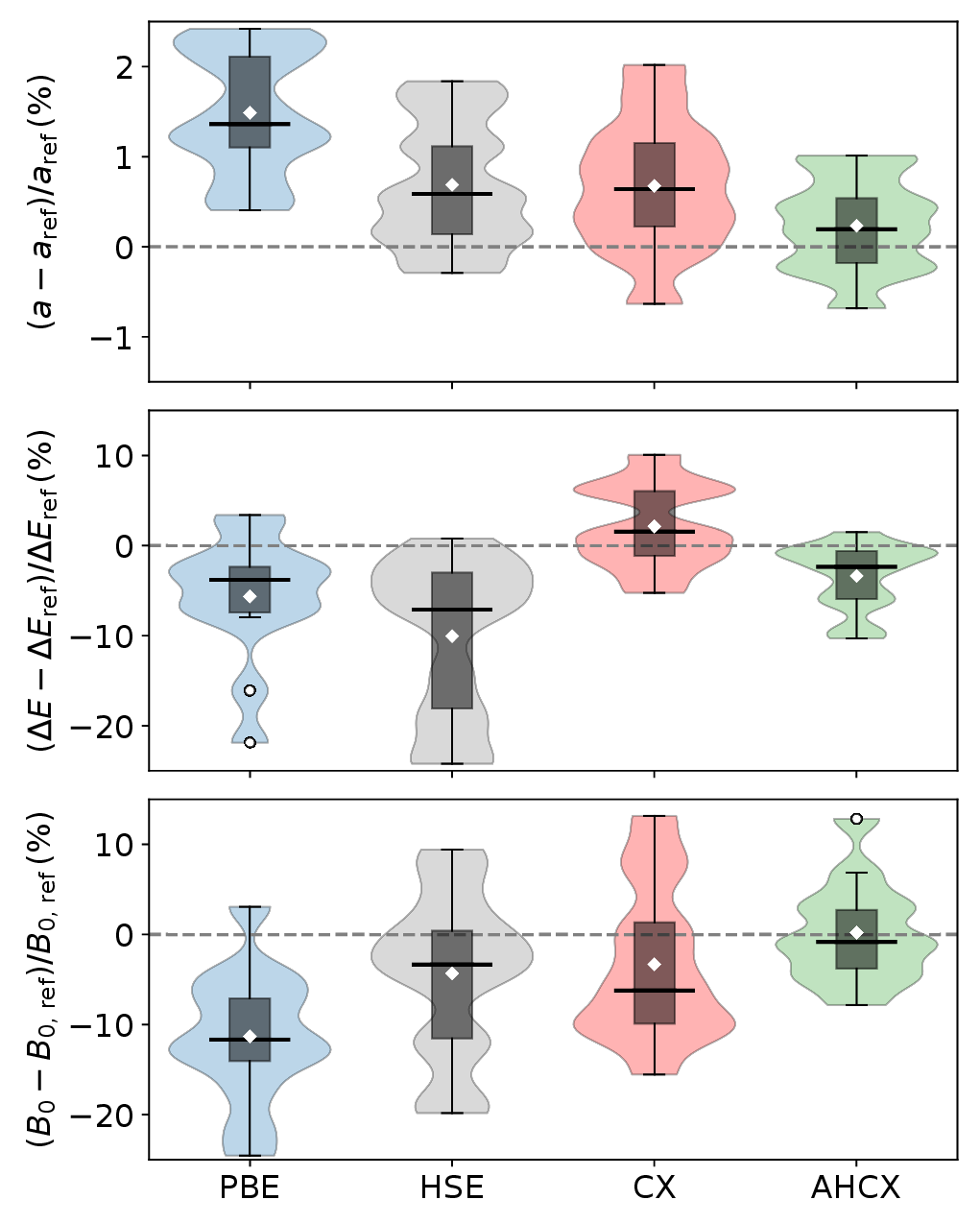} 
   \caption{Violin plot representing statistics of relative deviations in PBE, HSE, CX, and AHCX determinations of bulk lattice parameters (top), cohesive energies
   (middle), and bulk moduli (bottom panel). We compare with experimental values that are back-corrected for zero-point energy and thermal vibrational effects;  Details on the performance for each of the 13 non-magnetic elements and compounds (1 simple and 5 transition metals, 4 semiconductors and 3 ionic insulators) are given in SI material. Parts of the transition-metal performance comparison are detailed and further discussed in Table \ref{tab:oncv_bulk}. In the violin plots, the white diamonds mark the mean deviation, the box identifies the range between the first and third quartile, and the black line indicates the median of the distribution. Moreover, whiskers indicate the range of data falling within 1.5*box-lengths of the box.}
   \label{fig:violin_bulk}
\end{figure}

Figure \ref{fig:violin_bulk} contrasts the performance of hybrid functionals HSE and AHCX with those of the underlying PBE and CX functionals for bulk: 6 metals (elements Al, Cu, Rh, Ag, Pt, Au), 4 semiconductors (Si, C, SiC, and GaAs) and 3 ionic insulators (MgO, LiF, and NaCl). Here we compare deviations from measurements of lattice constants $a$ and cohesive energies $\Delta E$ (both back corrected for zero-point energy and thermal effects) as well as for bulk moduli $B_0$. Symbols for the statistical measures are summarized in the figure caption. The PBE (AHCX) description for cohesive energies (for bulk moduli) have Ag and Au (Rh) as statistical outliers; The SI material contains a detailed presentation of the per-bulk-material performance. 

The figure shows that AHCX gives the smallest median deviation and spread of data for the deviations for lattice parameters and bulk moduli; There are clear improvements 
with AHCX over both HSE and CX (which is the second-best performer overall). For cohesive energies, the CX has the smallest median deviation whereas AHCX has the smallest spread and both perform better than PBE and HSE for bulk
properties.

\begin{table}[h]
\caption{\label{tab:oncv_bulk} Performance assessment on bulk properties (lattice constants $a$ and cohesive energies $\Delta E$) for a set of late transition metals including the noble metals. The boldface entries are back-corrected experimental values \cite{Tran19}, the rest of the entries were obtained using, PBE, CX, and AHCX with ONCV PBE-SG15 PPs at $160$ Ry (except when labeled `*' where instead AbInit PPs was used at 80 Ry). Units are {\AA} for lattice parameters and eV for cohesive energies.}
\begin{ruledtabular}
\begin{tabular}{l|ccccc}
	System 		& $a_{\rm PBE}$ & $a_{\rm CX}$  &  $a_{\rm AHCX}$ 		& $a_{\rm AHCX}^{*}$    		    & \boldmath$a_{\rm ref}$     \\
\hline 
\textbf{Cu} 	& $3.639$ & $3.576$ & $3.587$                	 	& $3.591$              			    & $\mathbf{3.599}$ 	\\
\textbf{Ag} 	& $4.156$ & $4.065$ & $4.078$                		& $4.116$              			    & $\mathbf{4.070}$ 	\\
\textbf{Au} 	& $4.165$ & $4.101$ & $4.098$                		& $4.113$              			    & $\mathbf{4.067}$ 	\\
\textbf{Pt} 	& $3.970$ & $3.929$ & $3.910$                		& $3.939$              			    & $\mathbf{3.917}$ 	\\
\textbf{Rh} 	& $3.832$ & $3.786$ & $3.760$                		& $3.797$              			    & $\mathbf{3.786}$ 	\\
\hline \\[-2mm]
		&  $\Delta {E}_{\rm PBE}$ 
		&  $\Delta {E}_{\rm CX}$ 
		&  $\Delta {E}_{\rm AHCX}$ & $\Delta {E}_{\rm AHCX}^{*}$  & \boldmath$\Delta E_{\rm ref}$	\\
		\hline
\textbf{Cu} 	& $3.423$ & $3.781$ &  $3.348$             		 	& $3.765$                 		    & $\mathbf{3.513}$ 	\\
\textbf{Ag} 	& $2.488$ & $2.955$ &  $2.774$             		 	& $2.934$                 		    & $\mathbf{2.964}$  \\
\textbf{Au} 	& $2.997$ & $3.634$ &  $3.440$             		 	& $3.752$                 		    & $\mathbf{3.835}$  \\
\textbf{Pt} 	& $5.434$ & $6.226$ &  $5.524$             		 	& $4.876$                 		    & $\mathbf{5.866}$  \\
\textbf{Rh} 	& $5.565$ & $6.367$ &  $5.244$             		 	& $4.257$                 		    & $\mathbf{5.783}$  \\
\end{tabular}
\end{ruledtabular}
\end{table}

Table \ref{tab:oncv_bulk}
compares the results of PBE, CX, and AHCX 
descriptions of transition-metal structure and
cohesion to back-corrected experiments values\cite{Tran19} (last column). The results are provided for the 
electron-rich ONCV-SG15 PPs (except where noted by an
asterisk `*' or when taken from literature). 

We first observe that use of ONCV-SG15 PP setup yields a state-of-the-art 
benchmarking on these transition metal systems. For PBE (and for CX)
we can track deviations of our self-consistent ONCV results relative 
to the self-consistent PBE (non-selfconsistent CX) all-electron 
results, provided in Ref.\ \onlinecite{Tran19}, and to the fully self-consistent PAW-based assessments that one of us have previously provided.\cite{Gharaee2017} For lattice constants the largest PBE (CX) deviation from self-consistent (non-selfconsistent)
all-electron results\cite{Tran19} is 0.2\% (0.1\%). The ONCV characterization has but minute differences from (is spot on) from the PAW-based characterization of PBE 
(CX) performance.\cite{Gharaee2017} Similarly,
for the description of cohesive energies, we find  small PBE (CX) deviations, with a Rh maximum of 3.0\% (1.4\%), 
relative to all-electron results.\cite{Tran19} Here the ONCV characterizations are slightly less precise overall than the previous PAW-based characterizations 
(but have a smaller spread). Finally, 
for the bulk moduli, we find that our ONCV-SG15 benchmarking
differs by at most 2.2\% for Rh from the all-electron
transition-metal results. Overall, we find that
our ONCV-SG15 bulk-structure benchmarking is highly precise, 
for example, for characterizations of noble-metal properties.

Comparing next the calculated results
against back-corrected experimental results,\cite{Tran19} Table \ref{tab:oncv_bulk} confirms the overall impression
of AHCX promise (Fig.\ \ref{fig:violin_bulk}) also holds 
for the transition metals. The table shows
that CX performs significantly better than PBE on the 
transition metal lattice constants (cohesive energies); This
is also expected, perhaps especially for the noble 
metals.\cite{AmbSil16,Gharaee2017,Tran19,HyJiSh20}
A move from PBE to CX reduces the maximum absolute deviation (from experimental values) on transition-metal lattice constants (cohesive energies) from 2.4\% (24.2\%) to 0.8\% (10.1\%).  Meanwhile, AHCX performance is on par with that of CX for lattice constants and a clear improvement over HSE 
(as detailed in the SI material). This pattern is repeated for the bulk-modulus assessments, see SI material. 

Overall for AHCX we find the following deviations from experimental lattice-constant (cohesive-energy) values:\cite{Tran19} -0.3\% (-4.7\%) for Cu, 0.2\% (-6.4\%) for Ag, 0.8\% (-10.3\%) for Au, -0.2\% (-5.8\%) for Pt, and -0.7\% (-9.3\%) for Rh. The performance for cohesive
energies is overall slightly worse than that for CX on transition metals, but significantly better than that of PBE and HSE, see SI material. While broader tests of bulk-transition-metal performance of AHCX, for example, extending Refs.\ \onlinecite{Gharaee2017,Tran19} are 
desirable, it is also clear that AHCX is useful for descriptions of 
bulk properties in general and certainly for noble-metal studies.

\begin{table*}
\caption{\label{tab:NobleSurface} Results for workfunctions $\phi$ and surface energies $\sigma$  of 
the (111) facet of the noble metals. We compare PBE, CX, and AHCX against measurements -- in the case of surface energies using the procedure described in the text.  Calculations performed with the normconserving AbInit PPs at 80 Ry are marked by an asterisk `*', the rest are obtained using the ONCV-SG15 PPs at 160 Ry. All of the underlying slab calculations are done with surface relaxations, the primes indicate that those AHCX results are obtained at fully relaxed CX geometries (using bulk lattice constants obtained for CX with the AbInit PPs: 3.626 {\AA} for Cu, 4.152 {\AA} for Ag, and 4.146 {\AA} for Au).}
\begin{ruledtabular}
\begin{tabular}{llccccccl}
&	Property & PBE$^*$ & PBE & SCAN$^a$ & CX$^*$ & CX & AHCX$^*$ & Exp.\\
  \hline
Cu (111)         & $\phi$ [eV] &   4.81 & 4.80 & 4.98  & 4.96 & 5.00  & 5.05' 
         & 4.9$\pm$0.04$^{b,c}$  \\
         & $\sigma$ [j/m$^2$] &   1.26 & 1.32 & 1.49  & 1.71 & 1.81 & 1.66' & 1.76$\pm$0.18 \\
\hline
Ag (111)        & $\phi$ [eV] &  4.43 & 4.44 & 4.57  & 4.64 &  4.66 & 4.71'& 4.75$\pm$0.01$^{b,d}$ \\
         & $\sigma$ [j/m$^2$] &   0.69 & 0.75 & 0.97  & 1.07 & 1.18 & 1.18' & 1.17$\pm$0.12 \\
\hline
Au (111)        & $\phi$ [eV] &  5.22 & 5.20 & 5.32  & 5.39 & 5.39 & 5.57' & 5.3-5.6$^{b,e}$\\
         & $\sigma$ [j/m$^2$] &   0.72 & 0.73 & 0.93  & 1.17 & 1.21 &  1.28' & 1.38$\pm$0.14 \\
         \hline
         $^a$Ref.\ \onlinecite{PaBaJi20}. \\ 
         $^b$Ref.\ \onlinecite{derry2015recommended}. \\
         $^c$Ref.\ \onlinecite{haas1977work}. \\
         $^f$Ref.\ \onlinecite{farnsworth1940photoelectric}.\\
         $^e$Ref.\ \onlinecite{pescia1982spin}.\\
\end{tabular}
\end{ruledtabular}
\end{table*}

\subsection{Noble metal surfaces}

Descriptions of (organic) molecule adsorption at metal surfaces are 
an important reason for defining and launching the AHCX. For metal problems, we must screen the long-range Fock-exchange component. The new RSH vdW-DF is set up to reflect the perfect electrostatic screening.\cite{OTRSHalga}
Below we begin the discussion by looking 
at the AHCX ability, as a RSH vdW-DF, to describe properties of noble-metal surfaces. We stay with the robust (but expensive) electron-rich ONCV-SG15 PPs (at 160 Ry) for the regular-functional descriptions and for the description of CO adsorption, next subsection; For a demonstration of AHCX's ability to describe workfunctions and surface energies, here, we shall use and discuss calculations obtained in the more electron-sparse (but still normconserving) Ab Init PPs (at 80 Ry).

We note that understanding surface energies of metallic nanoparticles is important for correctly setting up cost-effective catalysis usage in catalysis. This is because they define the Wulff shapes\cite{Wulff} of nanoparticles and hence the extent that they provide access to specific active sites.\cite{LoRaBr20,Ageo20} As such this additional illustration could also be relevant for DFT practitioners.

Table \ref{tab:NobleSurface} compares our 
PBE, CX, and AHCX results for workfunctions and surface energies against experimental values and against literature SCAN values.\cite{PaBaJi20} An asterisk `*' on a column indicates that those results
are obtained using the AbInit PPs at 80 Ry, the rest are obtained by ONCV-SG15 PPs at 160 Ry. All results are 
based on a sequence of slab calculations and we first
compute bulk lattice constants and then surface relaxations
specific to the functional and PP choice. 
As indicated 
by the prime on the results in the AHCX column, however, we rely on the CX structure 
determination for our characterization of this CX-based RSH.

The table furthermore compares against experimental values for workfunctions (as summarized in 
Ref.\ \onlinecite{PaBaJi20}), and surface energies for noble-metal (111) facets.\cite{HyJiSh20,PaBaJi20}
Experimental surface-energy values are available from observations of the liquid metal surface tension.\cite{tyson1975surface,tyson1977surface} As such, they are defined by an average and it is natural to focus on the major facets\cite{PaBaJi20}
\begin{eqnarray}
    \sigma_{\rm exp} & = &  \sigma_{\rm avg} \, , \\
    \sigma_{avg} & = & \frac{1}{3} (\sigma_{(111)} + \sigma_{(110)} + \sigma_{(100)} ) \nonumber \\
    & \equiv & \frac{1}{3}\sigma_{(111)} + \frac{2}{3}\sigma_{\rm (other)} \, ,
    \label{eq:SurfAvg}
\end{eqnarray}
where we now introduce $\sigma_{\rm (other)}= (\sigma_{(110)} + \sigma_{(100)})/2$.  
Previous works have compared measured surface energies $\sigma_{\rm exp}$ (for various transition metals) and compared with per-facet surface-energy results -- 
$\sigma_{(111)}$, 
$\sigma_{(110)}$, 
and $\sigma_{(100)}$ -- computed in various functionals, including CX.\cite{PaBaJi20,HyJiSh20,LoRaBr20,Ageo20} 
Here, we simplify the assessment task in two steps: a) we use our past CX results\cite{HyJiSh20} to define a ratio
$x=\sigma_{(111)}/\sigma_{\rm (other)}$ that reflects the relative weight of (111) in the major-facet averaging,\cite{PaBaJi20,HyJiSh20} Eq.\ (\ref{eq:SurfAvg}),
and b) we extract the experimentally-guided estimate
\begin{equation}
    \sigma_{(111)}^{\rm exp.} \approx x*\sigma_{\rm (other)}  = \frac{3x}{2+x}* \sigma_{\rm exp} \, .
    \label{eq:SurfEstimate111}
\end{equation}

Table \ref{tab:NobleSurface} shows that moving between PBE, SCAN (as asserted in literature\cite{PaBaJi20}),
CX, and AHCX generally increases the computed values for both workfunctions and surface energies for the noble-metal (111) facets; We note that the same trend also arise (among the PBE, CX, and AHCX descriptions) when we instead compute these with the Ab Init PPs at 80 Ry. It is alone for the Cu surface energy that we find an AHCX surface energy that is smaller than the CX results (AHCX gives the same Ag surface energy as CX).

Table \ref{tab:NobleSurface} also shows that switching between the electron
sparse AbInit PPs and the electron-rich ONCV-SG15 PPs has essentially no impact on the workfunction and a 5\% (10\%) effect on the CX results for the 
surface energy of Cu and Au (of Ag). These shifts are directly correlated with the fact that the Cu and Au (Ag) lattice constant in the AbInit PP description is 1\% (2\%) larger than what results with the ONCV-SG15 PP choice.

Table \ref{tab:NobleSurface} suggests that both CX and AHCX
provide an accurate description of the noble-metal surface energies and workfunctions. We note that a significant uncertainty exists for surface energy reference values due to the need to interpret the input from high-temperature
measurements. Unlike for PBE and SCAN, the CX and AHCX
surface energies fall within the range of such observation-based estimates (For Au(111), the CX, but not the AHCX, result falls just outside the error bars). 
For workfunctions we find that CX and SCAN are slightly more accurate than AHCX for Cu(111), while AHCX is the best performer for the Ag(111) surface; 
For the Au(111) workfunction the SCAN, CX, and AHCX results all fall with the broad range of experimental values.

To argue AHCX accuracy for clean-noble-metal surface properties, we make four observations. First, our use of AbInit PPs for AHCX likely impacts the comparison with the ONCV-SG15 based PBE and CX results for the surface energy $\sigma$. Second the trend from the PBE and CX results suggests that the impact on the PP choice on $\sigma$
primarily arises by the inverse dependence on the surface-unit-cell area, an area-scaling that is discussed two paragraphs above. Third, Appendix A assesses the sensitivity of the CO-adsorption site-preference results to the PP choice, documenting good robustness for the noble metals;
This robustness is expected by the fact that these systems have a nearly closed $d$ band\cite{KresseNJP,HaNo95,Noblest}
and it makes it likely that the PP impact on the AHCX $\sigma$ description will be dominated by the area-scaling effect. Fourth, adjusting our AbInit-PP based AHCX characterization upwards by a 5-10\% area effect will generally push the AHCX description closer to the back-corrected experimental surface-energy values: The AHCX result for Ag will still reside on the edge of the range characterizing the Ag experimental data. In other words, Table \ref{tab:NobleSurface} shows that AHCX is useful and that it  may well be a strong performer for characterizations of clean-noble-metal surface properties.

\subsection{Noble metal adsorption}

\begin{figure}[t]
   \centering
   \includegraphics[width=1\linewidth]{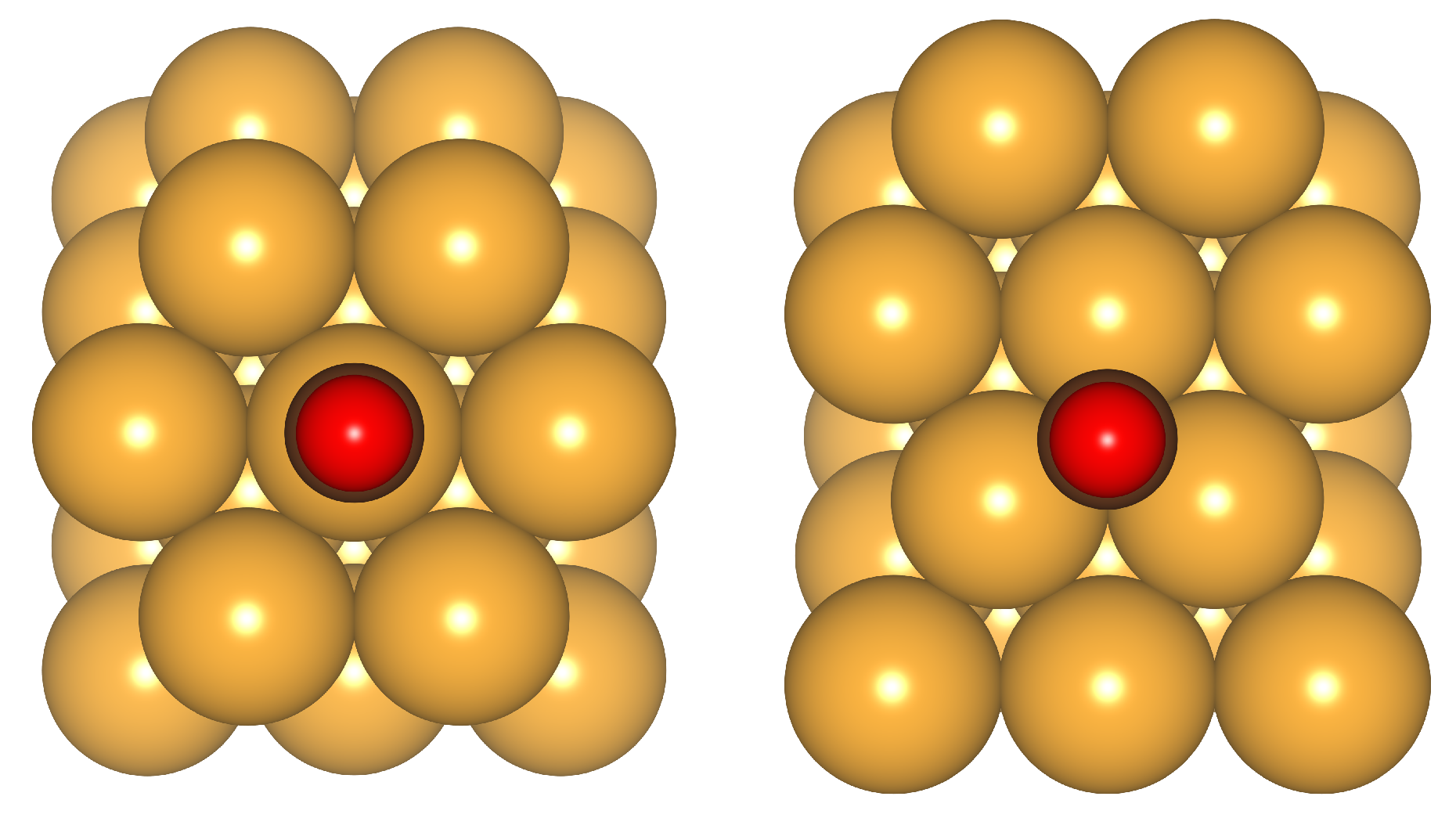}
   \caption{Schematics of CO adsorption on TOP and FCC
   (hollow) site on the fcc noble metals. Golden balls 
   show the surface and subsurface metal atoms, while the 
   red ball shows the oxygen position. The CO molecule is
   standing upright with the carbon atom (brown ball) sitting between the oxygen and the surface.}
   \label{fig:COadsorb}
\end{figure}

Finally, we turn to discuss a case of actual metal-surface adsorption. We focus on CO adsorption as it allows
us to use just a 2x2 in-surface unit-cell extension
and because it may provide a hint of whether AHCX 
ultimately may help in understanding 
heterogeneous catalysis. The CO adsorption on
late transition metals is a classic surface-science
challenge. 

Figure \ref{fig:COadsorb} shows the CO adsorption geometry for 111 surfaces. The CO is standing upright with a Blyholder-type binding\cite{Blyholder} through the carbon atoms (brown ball) which sits between the metal substrate (gold atoms) and the oxygen (red ball). There is no experimental ambiguity that CO molecule adsorbs on the TOP site (as illustrated in the right panel), rather than at the FCC or hollow site (left panel) at low coverage, Ref.\ \onlinecite{MYSHLYAVTSEV201551}.   

\begin{table}
\caption{\label{tab:Native_latt_Adsorb} Comparison of adsorption energies (eV) for CO on noble metals at TOP and FCC sites of the (111) facet,
calculated for PBE, CX and AHCX. Boldface entries identify a clear finding of the correct site preference. The AHCX results are computed at the adsorption geometry computed in CX, as indicated by a prime.}
\begin{ruledtabular}
\begin{tabular}{lcccccc}
         & Site &   PBE   & SCAN$^{a}$ & CX & AHCX' & Exp.\\
  \hline
Cu(111)  & FCC &   -0.829 & -1.01 & -0.964    & -0.695 & -- \\
         & TOP &   -0.718 & -0.88 & -0.850    & \bf{-0.719} & \bf{-0.50}$^{b}$\\
\hline
Ag(111)  & FCC &   -0.103  & -0.21 & -0.292   & -0.051 & --- \\
         & TOP &    \bf{-0.181} & -0.21 & \bf{-0.336} & \bf{-0.230} & \bf{-0.28}$^{c}$  \\
\hline
Au(111)  & FCC &   -0.212 & -0.45 & -0.465   & -0.197 & -- \\
         & TOP & -0.223 & -0.42 & -0.436   & \bf{-0.353}  & \bf{-0.40}$^{d}$\\

\hline
$^a$Ref.\ \onlinecite{patra2019rethinking} \\ 
$^b$Ref.\ \onlinecite{kessler1977chemisorption} \\ 
$^c$Ref.\ \onlinecite{mcelhiney1976adsorption} \\ $^d$Ref.\ \onlinecite{GSElliott1984}
\end{tabular}
\end{ruledtabular}
\end{table}

Table \ref{tab:Native_latt_Adsorb} summarizes the CO adsorption 
studies that we here provide using electron-rich ONCV-SG15 PPs. The most famous CO adsorption problem arises 
at the Pt(111) surface, Ref.\ \onlinecite{Feibelman01p4018,lazic10p045401,grinberg2002co,olsen2003co}, 
a problem that has repeatedly challenged DFT
descriptions when used for XC energy approximations 
that also deliver an accurate description of the in-surface 
Pt lattice constant and surface energies, 
Refs.\ \onlinecite{stroppa2007co,janthon2017adding,patra2019rethinking,alaei2008c,hammer1999improved}. Here we focus on the noble-metal adsorption cases. The table compares
our results for CO adsorption at FCC and TOP sites on Cu(111), Ag(111), and Au(111). As indicated by the prime, the AHCX characterization is again done frozen at the adsorption structure that results in the CX study. We find that the AHCX description stands out by predicting the correct site preference for CO on noble-metal (111) surfaces; None of the PBE, SCAN or CX functionals have that consistency. The 
AHCX prediction of the actual (TOP-site) adsorption energy is good in silver and 
gold; All functionals 
have some problems in accurately predicting the
magnitude of the adsorption energy for CO/Cu(111).

The AHCX functional overestimates the binding in the case of Cu(111). 
Comparing with the CX description, we see that the AHCX still provides a reduction of the adsorption energy. This AHCX binding softening (compared to CX) is, in part, expected.  For instance, use of either PBE0 and HSE reduce the Cu(111) adsorption energy compared with PBE, Ref.\ \onlinecite{stroppa2007co}. 

We also highlight an important point of the AHCX success: 
We find that AHCX  \textit{simultaneously} have accurate 
descriptions of molecules, bulk lattice constants, surface properties, and 
CO adsorption energies; There is no lucky hit for one quantity 
at the expense of others, compare 
Fig.\ \ref{fig:GMTKN53comp} as well as Tables \ref{tab:oncv_bulk}, 
\ref{tab:NobleSurface} and \ref{tab:Native_latt_Adsorb}. The AHCX 
lattice constant for Cu has a 0.3\% deviation from experiment,
better than CX (see SI material). The AHCX descriptions of 
workfunctions are accurate for Cu. Ag, and Au, and the description of Cu, Ag, and Au surface energies are good. Last but not least, the AHCX performance 
for predicting the CO-adsorption site preference is excellent, and (except for
Cu) accurate also on the predictions
of the actual CO adsorption energies.

\section{Summary and conclusion} 

We find  that the new AHCX RSH 
performs clearly  better than both CX and dispersion-corrected HSE for molecules and better than CX and HSE for extended matter. For molecules, our new AHCX RSH performs at the level of the corresponding unscreened CX0p hybrid.

Furthermore we find that the AHCX is accurate for the description of noble-metal surface properties, at least for the (111) facet.  The AHCX improves both the PBE and CX descriptions in terms of being overall more accurate on surface workfunctions and surface energies for the (111) noble-metal facets. It furthermore stands 
out by predicting the right site preference for CO 
adsorption on the (111) facet across the noble metals. For Ag and Au the AHCX adsorption energies are accurate, but the binding strength is overestimated for Cu(111). However, at a geometry fixed by the CX structure characterization, the AHCX still improves the CX description of the CO/Cu(111)
adsorption binding energy.

Overall the AHCX shows promise for tackling a long-standing problem of describing organics-metal interfaces. In the heterogeneous systems we often want a vdW-inclusive hybrid for the molecule side, yet we cannot motivate the use of unscreened hybrid CX0p in cases with a metallic nature of conduction for the substrate. The new screened CX-based RSH, termed AHCX, shows overall, a significantly better performance on molecules than dispersion-corrected HSE. Our results also indicate a strong performance on semiconducting and metallic systems. It is a candidate for making DFT better at characterizing chemistry in heterogeneous systems.

\section*{SUPPLEMENTARY MATERIAL}
See the supplementary material for the parameters of HJS analytical-hole, computational time scaling, molecular and bulk benchmark data. 

\section *{Acknowledgement}

We thank Paolo Giannozzi, Jung-Hoon Lee, Jeffrey B. Neaton, 
and Elsebeth Schr{\"o}der for useful discussions.
Work is supported by the Swedish Research 
Council (VR) through  Grants  No.\  2014-4310  and   
2018-03964, the Swedish Foundation for Strategic research (SSF) through grant IMF17-0324, Sweden's Innovation Agency (Vinnova) through project No.:2020-05179, as well as the Chalmers Area-of-Advance-Materials \& -Production theory activities. The authors also acknowledge computer allocations from the Swedish National Infrastructure for Computing (SNIC), under Contracts  SNIC2019-1-12, SNIC 2020-3-13,  and from the Chalmers Centre for Computing, Science and Engineering (C3SE).

\appendix

\begin{table}
\caption{\label{tab:PPsensitivity} Sensitivity of CO-adsorption results (listed in eV) for TOP and FCC sites of the (111) facet, comparing CX and AHCX in the AbInit PPs (as indicated by an asterisk on columns) with the set of main-paper ONCV-SG15 results (repeated for convenience). The AHCX results are provided at the adsorption geometry computed in CX (at the relevant PP choice), as indicated by primes.
}
\begin{ruledtabular}
\begin{tabular}{lccccc}
Metal-Site & CX$^*$ & CX   & AHCX$^*$' & AHCX' & Exp.\\
  \hline
Cu-FCC &   -0.964 & -1.118 & -0.800    & -0.695 & -- \\
Cu-TOP &   -0.850 & -0.946 & -0.841    & -0.719 & \bf{-0.50}$^{a}$\\
\hline
Ag-FCC &   -0.292 & -0.361 & -0.165    & -0.051 & --- \\
Ag-TOP &   -0.336 & -0.376 & -0.295    & -0.230 & \bf{-0.28}$^{b}$  \\
\hline
Au-FCC &   -0.465 & -0.514 & -0.326    & -0.197 & -- \\
Au-TOP &   -0.436 & -0.461 & -0.407    & -0.353  & \bf{-0.40}$^{c}$\\

\hline
$^a$Ref.\ \onlinecite{kessler1977chemisorption} \\ 
$^b$Ref.\ \onlinecite{mcelhiney1976adsorption} \\ 
$^c$Ref.\ \onlinecite{GSElliott1984}
\end{tabular}
\end{ruledtabular}
\end{table}

\section{Sensitivity to pseudopotentials}

Hybrid calculations are most stable for normconserving PPs in the version of 
\textsc{Quantum Espresso} that we used for our AHCX implementation and subsequently for our documentaion work; A workaround was introduced recently
in the \textsc{Quantum Espresso} code package, but it does not seem to
always help and, in any case, it did not arive in time for us to benefit in this 
project.  The stability concerns exist whether we use the existing HSE\cite{HSE03,HSE06} or our coding of analytical-hole formulations.\cite{HJS08}
Meanwhile, it is convenient to have a (nearly) complete suite for materials explorations. The AbInit PP\cite{abinit05} and the ONCV-SG15\cite{ONCV,sg15} PP releases are two such options.

We primarily use the electron-rich ONCV-SG15 because it included semi-core electrons as valence states to allow simpler discussion of AHCX performance examples. Use of the ONCV-SG15 is a costly option but also considered safe, for example, for hybrid studies.

The AbInit PPs\cite{abinit05} are also normconserving
but electron sparse, relying instead on a  nonlinear
core correction to represent the remainder.
One must expect 
a lower accuracy than what is available with 
ONCV-SG15\cite{ONCV,sg15} or from PAW-based
structure descriptions, e.g., Ref.\ \onlinecite{Gharaee2017}; Comparing
the AHCX columns with and without an 
asterisk `*' in Table \ref{tab:oncv_bulk} gives an illustration. Also, relying on a nonlinear core-corrections, the use of the
AbInit PPs can cause problems for hybrid descriptions of adsorption on open-$d$-shell systems.\cite{KresseNJP} However, since they need fewer band (lower the computation costs) and have memory requirements that fits more computer resources, the electron-sparse approach may sometimes be an interesting option. We used the AbInit PPs to provide an additional demonstration of the potential AHCX usefulness regarding noble-metal workfunctions and surface energies. We did that noting that we work with nearly-closed-shell systems (where the importance of $d$-band rehybridization is generally reduced\cite{HaNo95,Noblest,stroppa2007co}).

Table \ref{tab:PPsensitivity} reports an assessment of the impact of PP choices on CX and AHCX adsorption energy results for CO on noble-metal surfaces. Here and in a molecular
survey (not shown), we find that using instead the electron-sparse AbInit PPs introduces no systematic changes in the conclusion that AHCX is accurate for molecules and for predicting the site preference for noble-metal adsorption. The ONCV-SG15 descriptions are more accurate in terms of giving better lattice constants and better absolute adsorption values across the noble metals. However, we also find working descriptions with the computationally cheaper AbInit PPs, for noble metals.
This AHCX robustness on adsorption suggests that AHCX is, in fact, promising also on workfunction and surface energy results 
(the additional demonstrator included in Section V.C). A future ONCV-SG15 characterization of noble-metal surface energies 
and workfunctions may quantify the impact of this approximation, but some problems are simple enough that we can leverage the acceleration and memory savings, when desired.

%

\cleardoublepage
\pagebreak
\widetext
\begin{center}
\textbf{\large Supplementary Materials for:\\
vdW-DF-ahcx: a range-separated van der Waals density functional hybrid}
\end{center}
\setcounter{section}{0}
\setcounter{equation}{0}
\setcounter{figure}{0}
\setcounter{table}{0}
\setcounter{page}{1}
\makeatletter
\renewcommand{\theequation}{S \arabic{equation}}
\renewcommand{\thetable}{S \Roman{table}}
\renewcommand{\thefigure}{S \arabic{figure}}
\renewcommand{\bibnumfmt}[1]{[S#1]}


\section{HJS model of exchange holes: vdW-DF releases}
Table \ref{tab:paramHmore} lists the parameters of HJS analytical-hole exchange-hole descriptions for revPBEx (exchange  in vdW-DF1), rPW86 (vdW-DF2) and cx13 (exchange in CX). The parameters define the dependence with scaled density gradient $s$ of the  Gaussian-suppression functions $\mathcal{H}(s)$ that characterize the overall exchange-hole shape. 
The table supplements the comparison
between PBE, PBEsol, and CX 
exchange descriptions reported
in Table I and in Figs.\ 1 and 2 
of the main text.
\begin{table}[h]
\caption{\label{tab:paramHmore} Parameters of the rational function defining the Gaussian suppression $\mathcal{H}(s)$ in the HJS AH model for exchange in the vdW-DF releases. The revPBEx is used in vdW-DF1, the rPW86 in vdW-DF2, and the cx13 or LV-rPW86 in the CX.}
\begin{ruledtabular}
\begin{tabular*}{0.9\textwidth}{@{\extracolsep{\fill}}lrrr}
      &        revPBEx  &    rPW86  &      cx13 \\
\hline
$a_2$ &      0.0152730 &      0.0000006 &      0.0024387     \\
$a_3$ &     -0.0364003 &      0.0402647 &     -0.0041526       \\
$a_4$ &      0.0357444 &     -0.0353219 &      0.0025826        \\
$a_5$ &     -0.0092754 &      0.0116112 &      0.0000012       \\
$a_6$ &     -0.0098175 &     -0.0001555 &     -0.0007582        \\
$a_7$ &      0.0069143 &      0.0000504 &      0.0002764       \\
$b_1$ &     -2.8845683 &     -1.8779594 &     -2.2030319        \\
$b_2$ &      3.7112964 &      1.5198811 &      2.1759315        \\
$b_3$ &     -2.7291409 &     -0.5383109 &     -1.2997841       \\
$b_4$ &      1.2683681 &      0.1352399 &      0.5347267       \\
$b_5$ &     -0.3661883 &     -0.0428465 &     -0.1588798       \\
$b_6$ &      0.0465588 &      0.0117903 &      0.0367329        \\
$b_7$ &      0.0126146 &      0.0033791 &     -0.0077318        \\
$b_8$ &     -0.0035666 &     -0.0000493 &      0.0012667       \\
$b_9$ &      0.0028382 &      0.0000071 &      0.0000008        \\
\end{tabular*}
\end{ruledtabular}
\end{table}

\section{Computational time scaling}
\begin{table}[b]
\begin{ruledtabular}
\begin{center}
\small
\caption{Test of scaling in computational cost (cpu hour) and convergence for HSE, CX0p and the AHCX. The data is plotted in Fig.\ 3 of the main text.}
\label{tbl:example}
\begin{tabular}{c c | r r | r r}
         & &  \multicolumn{2}{c}{\textbf{Au}}  &  \multicolumn{2}{c}{\textbf{C}} \\
         &q mesh   & time/FX &  $E_{\rm coh}$ & time/FX &  $E_{\rm coh}$\\
\hline
HSE06   & 8x8x8 & 25.26 & -2.948 &  1.35 & -7.510 \\
        & 4x4x4 &  3.15 & -2.900 &  0.18 & -7.491 \\
        & 2x2x2 &  0.42 & -3.064 &  0.03 & -7.558 \\
CX0p    & 8x8x8 &  25.52 & -3.477 & 1.36  & -7.570 \\
        & 4x4x4 &   3.23 & -3.497 & 0.18 & -7.570 \\
        & 2x2x2 &   0.42 & -3.666 & 0.03 & -7.628 \\
AHCX    & 8x8x8 &  25.26 & -3.487 &  1.35 &  -7.565  \\
        & 4x4x4 &   3.15 & -3.448 &  0.18 & -7.550 \\
        & 2x2x2 &   0.42 & -3.581 &  0.03 & -7.603 \\
\end{tabular}
\label{tab:scaling}\end{center}
\end{ruledtabular}
\end{table}
Table \ref {tbl:example} shows the time for converging one complete Fock-exchange evaluation step in the \textsc{Quantum ESPRESSO} code with  hybrid functionals HSE, vdW-DF-CX0p (CX0p,) and vdW-DF-AHCX (AHCX). The calculations of cohesive energies
$E_{\rm coh}$ were done at the (back-corrected) experimental lattice parameters of gold and diamond to allow the reader to also track the change in evaluation precision with the $q$ 
mesh of $k$-point differences in the Fock exchange evaluation.
We used an $8\times8\times$ $k$ point sampling, the 
ONCV-sg15\cite{ONCV,sg15} set of normconserving 
pseudopotentials and a 160 Ry wavefunction energy cut off. 
We also used 2 extra bands for the semiconductor and 10 extra bands for the metal cases and note that (extra) bands cost
in hybrid calculations.\cite{PaoloElStruct1} Besides the resulting $E_{\rm coh}$ description, we list the average central-processor-unit(cpu) core-hour cost per Fock-exchange evaluation step for various choices of $q$ meshes.

The timing results reflect calculations on 1 node (20 cores
sharing 60 GB of memory) of a high-performing computer cluster. The data is plotted in Fig.\ 3 of the main text. The listed cohesive energies should not be considered as more than an indication of functional performance as they were 
not done at the lattice constant native to our functional. Instead, we refer to Table III, section V.B, and the appendix in the main text, as well as to SI Tables \ref{tab:ecoh} and \ref{tab:lattice} (below,) for a more complete 
assessment of AHCX performance on bulk properties.

\section{Molecular benchmarks}
\vspace{-1cm}
\begin{table}[h]
\caption{ Performance of vdW-DFs and of a few vdW-inclusive DFTs
as asserted on the intermolecular and intramolecular NCI
benchmark groups of the GMTKN55 suite.\cite{gmtkn55} 
As indicated by the asteriks `*', we exclude the WATER27 
benchmark set in the intermolecular NCI group. We list
so-called TMAD values in kcal/mol. The TMAD values are defined 
in the text and reflect mean absolute deviation 
(MAD) values averaged within the two NCI benchmark groups. 
\label{tab:NCIquantified}
}
\begin{ruledtabular}
\begin{tabular*}{0.48\textwidth}{@{\extracolsep{\fill}}lrr}
      &   Intermol. NCI$^*$ &   Intramol. NCI \\
\hline
vdW-DF1      &    0.92       &      1.36       \\
vdW-DF2      &    0.68       &      1.20       \\
rVV10        &    0.74       &      0.90       \\
HSE-D3       &    0.69       &      0.72     \\
rev-D3       &    0.58       &      0.75       \\
vdW-DF-ob86  &    0.55       &      0.72       \\
vdW-DF2-b86r &    0.46       &      0.65       \\
CX           &    0.47       &      0.63       \\
CX0p         &    0.45       &      0.49       \\
AHCX         &    0.44       &      0.45       \\
\end{tabular*}
\end{ruledtabular}
\end{table}
Table \ref{tab:NCIquantified} 
lists the data plotted in Fig.\ 5 in the main text. We track the performance across essentially the full set of noncovalent interaction (NCI) benchmarks of the GMTKN55 suite. We exclude
the WATER27 set for reasons explained in the main text; This 
also implies that our vdW-DF assessments, plotted in Fig. 5 of the main text, can be directly compared with a literature survey of dispersion-corrected meta-GGA performance, Ref.\ \onlinecite{Jana20}.

\begin{equation}
\rm{Intermolecular TMAD} = \frac{\sum_i MAD_i}{N} \, .
\end{equation}

Specifically, the data in NCI sets is resolved into and analyzed in terms of a total mean-average-deviation (MAD) values 
(denoted TMAD) for GMTKN55 groups 4 (intermolecular NCI) and for group 5 (intramolecular NCI). These measures arise by computing the MAD$_i$ (and MAD$_j$) values for every NCI benchmark set $i$ in group 4 (and $j$ in group 5), and then averaging over the 
number $N=11$ (and $M=9$) of sets in the group, for example,  

\begin{table*}[t]
\caption{\label{tab:GMTKN53_wtmad} Weighted total mean absolute deviation assessments (termed WTMAD1 and WTMAD2)
of functional performance on broad molecular properties; See also Fig.\ 6 of the paper. Our assessment is defined by the GMTKN55 suite \cite{gmtkn55} but we focus on the 53 sets that are easily accessible to our planewave benchmarking using the electron-rich ONCV-sg15 pseudopotentials at 160 Ry wavefunction energy cut off.
Use of the WTMAD1 and WTMAD2 measures (adapted to averaging over 53 sets) permits a meaningful performance comparison both within and among the benchmark Groups 1-6 that are also defined in Ref.\ \onlinecite{gmtkn55}.
As indicated by an astriks '*', we omit the G21EA sets in Group 1 and the WATER27 set in Group 4. For reference we also compare the performance of the CX against
revPBE-D3, a strong performing dispersion-corrected GGA. \cite{gmtkn55} Best performance WTMAD1 numbers (as asserted in the 53 benchmark subset) for total-noncovalent interactions (Group 6) and overall are highlighted.} 
\begin{ruledtabular}
\begin{tabular}{lcccccccc}
Functional  &  Measure & Group 1$^*$   & Group 2 &   Group 3    &  Group 4$^*$   & Group5 &   Group 6   &  GMTKN53     \\
\hline
revPBE-D3   & WTMAD1 & 5.04 &      5.55 &      6.30 &      3.69 &      4.27 &      3.95 &   4.88\\
            & WTMAD2 & 5.87 &     10.30 &     14.88 &      6.80 &      8.08 &      7.46 &   8.43\\
HSE-D3      & WTMAD1 & 3.98 &      4.42 &      4.47 &      4.40 &      4.26 &      4.33 &     4.25\\
            & WTMAD2 & 4.66 &      8.68 &      9.10 &      7.58 &      7.98 &      7.78 &     7.15\\
CX          & WTMAD1 & 5.07 &      4.99 &      7.56 &      3.44 &      4.04 &      3.71 &     4.87\\
            & WTMAD2 & 6.26 &      9.60 &     18.23 &      8.40 &      7.37 &      7.87 &     9.05\\
CX0p        & WTMAD1 & 3.77 &      3.61 &      4.52 &      3.34 &      3.06 &      3.22 &    3.63\\
            & WTMAD2 & 4.50 &      6.24 &     10.40 &      7.91 &      5.58 &      6.71 &     6.44\\
AHCX        & WTMAD1 & 3.77 &      3.64 &      4.54 &      3.22 &      2.93 &      \textbf{3.09} &     \textbf{3.59}\\
            & WTMAD2 & 4.51 &      6.25 &     10.47 &      7.66 &      5.38 &      6.49 &     6.37\\
\end{tabular}
\end{ruledtabular}
\end{table*}
~                          

Table \ref{tab:GMTKN53_wtmad} summarizes the assessments we have performed for the 53-benchmark subset  of  the full GMTKN55 suite.\cite{gmtkn55} This data is organized into the benchmark groups of the GMTKN55 suite, omitting, however, the WATER27 set in group 4 and the G21EA set in Group 1, again see the main text. The data is plotted in Fig.\ 6.

We list the weighted total mean absolute deviation measures (WTMAD1 and WTMAD2), essentially as  introduced by Grimme and co-workers in Ref.\ \onlinecite{gmtkn55}.
However, we have adapted the measures to the slightly smaller range of benchmark sets, for example, 
\begin{equation}
    \hbox{\rm WTMAD1'}  = \left(\sum^{53}_{i=1} w_i \times {\rm MAD}_i\right)/53 \, ,
    \label{eq:WTMAD1}
\end{equation}
Here, as in Ref.\ \onlinecite{gmtkn55}, 
we weight contributions of set $i$
according to the reference value
$\overline{|\Delta E|}_i$. The weighting $w_i$ is set to 10 (0.1) if $\overline{|\Delta E|}_i$
is smaller than 7.5 kcal/mol 
(larger than 75 kcal/mol); The weighting
$w_i$ is set to unity otherwise.

For the alternative WTMAD2 scheme, one 
relies on a the overall suite average
absolute deviation energy, $\overline{|\Delta E|}_{\rm suite}=56.84$ kcal/mol. It is computed from reference energies by considering all the reactions that enters in the GMTKN55 suite. 
Adapting to the reduced number (53) of 
sampled benchmark sets, we report
and compare per-functional values 
\begin{equation}
    \hbox{\rm WTMAD2'} =  \left(\sum\limits^{53}_{i=1} N_i  \frac{\overline{|\Delta E]|}_{\rm suite}}{\overline{|\Delta E|}_i} 
    \times {\rm MAD}_i \right)/\sum\limits^{53}_{i=1}N_i \, ,
    \label{eq:WTMAD2}
\end{equation}
where $N_i$ denotes the number of reactions
in the individual benchmark set $i$.

With a small further adjustment, we can
also use Eq. (1) and (2) to provide
a per-group assessment of functional
performance, Table \ref{tab:GMTKN53_wtmad} and Fig.\ 6. For example, for WTMAD1, we simply a) limit the summations over $i$ to the specific benchmark sets entering in each group and b) adjust the denominator 
to the number of benchmark sets that 
enters in the specific group. 

While our WTMAD' forms, Eqs (1) and (2), are slightly adjusted, we still
discuss them and compare them as 
estimates of full-suite WTMAD values.\cite{gmtkn55} This is done, for example, in Fig.\ 6 of the main text (while also using an asteriks `*' to emphasize the adjustment). Comparisons are motivated because we have merely removed one set out of 11 in Group 4, and 1 sets out of 18 in Group 1, omissions that can only
make small changes in  averaged assessments. In fact, we have validated 
that the weighted MAD values reported for AHCX in Table \ref{tab:GMTKN53_wtmad} \textit{undersells} our new hybrid vdW-DF performance
by 0.2 kcal/mol on the WTMAD1 measures and by 0.4 kcal/mol WTMAD2 
measure when done at the level of electron-rich ONCV-sg15 PPs.\cite{ONCV,sg15}
To that end, we obtained characterization in AHCX also for 
the WATER27 and G21EA sets that carefully controlled the impact of the
the self-interaction errors,\cite{Burke} as will be published elsewhere. 
\clearpage
\section{Bulk properties}
Tables \ref{tab:lattice}, \ref{tab:ecoh} and \ref{tab:ecoh} list the per-system 
deviations (relative to back-corrected experiments) on bulk 
lattice-constant, cohesive-energy and bulk modulus results, i.e., the data 
used in the violin-plot Fig.\ ~7 in the main text. The results 
were obtained using  ONCV-sg15 pseudopotential set and 
160 Ry energy cut off. 
\begin{table}[h]
\caption{\label{tab:lattice} Performance assessment on lattice constants, for a set of simple metals, semiconductors and ionic compounds. The table compares 
results in {\AA} (computed using ONCV-sg15 pseudopotentials at $160$ Ry) with all-electron Wien2K results and back-corrected experimental values. The percentage deviation from the experimental values are given in parenthesis.}
\begin{ruledtabular}
\begin{tabular}{llllllll}
System        	& ${{\rm PBE}_{\rm W2k}}^{a}$ & ${{\rm CX}_{\rm W2k}}^{a}$  & ${\rm PBE}_{\rm ONCV}$ 	&  ${\rm HSE}_{\rm ONCV}$  &  ${\rm CX}_{\rm ONCV}$ & ${\rm AHCX}_{\rm ONCV}$ & ${\rm Exp}^{a}$ \\
\hline 
\textbf{Cu} 	&  $3.632(  0.9\%)$ & 	 $3.579(-0.6\%)$ & 	 $3.639(  1.1\%)$  & 	 $3.638(  1.1\%)$  & 	 $3.576( -0.6\%)$ & 	 $3.587( -0.3\%)$ & 	 $ 3.599$ \\
\textbf{Ag} 	&  $4.148(  1.9\%)$ & 	 $4.065(-0.1\%)$ & 	 $4.156(  2.1\%)$  & 	 $4.145(  1.8\%)$  & 	 $4.065( -0.1\%)$ & 	 $4.078(  0.2\%)$ & 	 $ 4.070$ \\
\textbf{Au} 	&  $4.161(  2.3\%)$ & 	 $4.095( 0.7\%)$ & 	 $4.165(  2.4\%)$  & 	 $4.129(  1.5\%)$  & 	 $4.101(  0.8\%)$ & 	 $4.098(  0.8\%)$ & 	 $ 4.067$ \\
\textbf{Pt} 	&  $3.971(  1.4\%)$ & 	 $3.927( 0.3\%)$ & 	 $3.970(  1.4\%)$  & 	 $3.921(  0.1\%)$  & 	 $3.929(  0.3\%)$ & 	 $3.910( -0.2\%)$ & 	 $ 3.917$ \\
\textbf{Rh} 	&  $3.832(  1.2\%)$ & 	 $3.789( 0.1\%)$ & 	 $3.832(  1.2\%)$  & 	 $3.779( -0.2\%)$  & 	 $3.786(  0.0\%)$ & 	 $3.760( -0.7\%)$ & 	 $ 3.786$ \\
\textbf{Al}     &  $4.041(  0.5\%)$ & 	 $4.029( 0.2\%)$ & 	 $4.044(  0.5\%)$  & 	 $4.039(  0.4\%)$  & 	 $4.041(  0.5\%)$ & 	 $4.033(  0.3\%)$ & 	 $ 4.022$ \\
\textbf{Si}     &  $5.471(  1.1\%)$ & 	 $5.441( 0.5\%)$ & 	 $5.477(  1.2\%)$  & 	 $5.446(  0.6\%)$  & 	 $5.462(  0.9\%)$ & 	 $5.441(  0.5\%)$ & 	 $ 5.412$ \\
\textbf{C}      &  $3.575(  0.6\%)$ & 	 $3.567( 0.4\%)$ & 	 $3.567(  0.4\%)$  & 	 $3.543( -0.3\%)$  & 	 $3.561(  0.2\%)$ & 	 $3.545( -0.2\%)$ & 	 $ 3.553$ \\
\textbf{SiC}    &  $4.385(  0.9\%)$ & 	 $4.369( 0.5\%)$ & 	 $4.381(  0.8\%)$  & 	 $4.352(  0.1\%)$  & 	 $4.374(  0.6\%)$ & 	 $4.353(  0.2\%)$ & 	 $ 4.346$ \\
\textbf{GaAs}   &  $5.749(  1.9\%)$ & 	 $5.680( 0.7\%)$ & 	 $5.751(  2.0\%)$  & 	 $5.652(  0.2\%)$  & 	 $5.705(  1.2\%)$ & 	 $5.640(  0.0\%)$ & 	 $ 5.640$ \\
\textbf{MgO}    &  $4.259(  1.7\%)$ & 	 $4.231( 1.0\%)$ & 	 $4.255(  1.6\%)$  & 	 $4.209(  0.5\%)$  & 	 $4.243(  1.3\%)$ & 	 $4.205(  0.4\%)$ & 	 $ 4.189$ \\
\textbf{LiF}    &  $4.070(  2.5\%)$ & 	 $4.056( 2.1\%)$ & 	 $4.062(  2.3\%)$  & 	 $4.016(  1.1\%)$  & 	 $4.052(  2.0\%)$ & 	 $4.012(  1.0\%)$ & 	 $ 3.972$ \\
\textbf{NaCl}   &  $5.700(  2.4\%)$ & 	 $5.661( 1.7\%)$ & 	 $5.698(  2.3\%)$  & 	 $5.663(  1.7\%)$  & 	 $5.661(  1.7\%)$ & 	 $5.623(  1.0\%)$ & 	 $ 5.569$ \\
\hline
\textbf{MD} 		 & $0.066$ 		 & $0.027$ 		 & $0.066$ 		 & $0.030$ 		 & $0.032$ 		 & $0.011$ 		 &  \\
\textbf{MAD} 		 & $0.066$ 		 & $0.031$ 		 & $0.066$ 		 & $0.033$ 		 & $0.036$ 		 & $0.019$ 		 &  \\
\textbf{RMSD} 		 & $0.074$ 		 & $0.041$ 		 & $0.074$ 		 & $0.043$ 		 & $0.046$ 		 & $0.024$ 		 &  \\
\hline
$^a$ Ref. \onlinecite{Tran19}
\end{tabular}
\end{ruledtabular}
\end{table}

\begin{table}[h]
\caption{\label{tab:ecoh} Performance assessment on cohesive energies, for a set of simple metals, semiconductors and ionic compounds. The table compares results in eV (computed using ONCV-sg15 pseudopotentials at $160$ Ry) with all-electron Wien2k results and back-corrected experimental values. The percentage deviation from the experimental values are given in parenthesis.} 
\begin{ruledtabular}
\begin{tabular}{llllllll}
System        	& ${{\rm PBE}_{\rm W2k}}^{a}$ & ${{\rm CX}_{\rm W2k}}^{a} $  & ${\rm PBE}_{\rm ONCV}$ 	&  ${\rm HSE}_{\rm ONCV}$  &  ${\rm CX}_{\rm ONCV}$ & ${\rm AHCX}_{\rm ONCV}$ & ${\rm Exp.}^{a}$ \\
\hline 
\textbf{Cu} 	&  $3.52(  0.2\%)$ & 	 $3.83( 9.0\%)$ & 	 $3.423( -2.6\%)$  & 	 $3.027(-13.8\%)$  & 	 $3.781(  7.6\%)$ & 	 $3.348( -4.7\%)$ & 	 $ 3.513$ \\
\textbf{Ag}     &  $2.53(-14.6\%)$ & 	 $2.99( 0.9\%)$ & 	 $2.488(-16.1\%)$  & 	 $2.368(-20.1\%)$  & 	 $2.955( -0.3\%)$ & 	 $2.774( -6.4\%)$ & 	 $ 2.964$ \\
\textbf{Au}     &  $3.03(-21.0\%)$ & 	 $3.64(-5.1\%)$ & 	 $2.997(-21.9\%)$  & 	 $2.917(-23.9\%)$  & 	 $3.634( -5.2\%)$ & 	 $3.440(-10.3\%)$ & 	 $ 3.835$ \\
\textbf{Pt}     &  $5.55( -5.4\%)$ & 	 $6.31( 7.6\%)$ & 	 $5.434( -7.4\%)$  & 	 $4.811(-18.0\%)$  & 	 $6.226(  6.1\%)$ & 	 $5.524( -5.8\%)$ & 	 $ 5.866$ \\
\textbf{Rh}     &  $5.74( -0.7\%)$ & 	 $6.46(11.7\%)$ & 	 $5.565( -3.8\%)$  & 	 $4.384(-24.2\%)$  & 	 $6.367( 10.1\%)$ & 	 $5.244( -9.3\%)$ & 	 $ 5.783$ \\
\textbf{Al}     &  $3.44(  0.3\%)$ & 	 $3.61( 5.2\%)$ & 	 $3.517(  2.5\%)$  & 	 $3.404( -0.8\%)$  & 	 $3.642(  6.1\%)$ & 	 $3.440( -0.3\%)$ & 	 $ 3.431$ \\
\textbf{Si}     &  $4.57( -2.5\%)$ & 	 $4.80( 2.5\%)$ & 	 $4.575( -2.3\%)$  & 	 $4.545( -3.0\%)$  & 	 $4.758(  1.6\%)$ & 	 $4.664( -0.4\%)$ & 	 $ 4.685$ \\
\textbf{C}      &  $7.71(  3.5\%)$ & 	 $7.79( 4.5\%)$ & 	 $7.705(  3.4\%)$  & 	 $7.511(  0.8\%)$  & 	 $7.891(  5.9\%)$ & 	 $7.565(  1.5\%)$ & 	 $ 7.452$ \\
\textbf{SiC}    &  $6.40( -1.2\%)$ & 	 $6.58( 1.6\%)$ & 	 $6.407( -1.1\%)$  & 	 $6.322( -2.4\%)$  & 	 $6.590(  1.7\%)$ & 	 $6.406( -1.1\%)$ & 	 $ 6.478$ \\
\textbf{GaAs}   &  $3.15( -5.6\%)$ & 	 $3.41( 2.2\%)$ & 	 $3.119( -6.5\%)$  & 	 $3.143( -5.8\%)$  & 	 $3.358(  0.6\%)$ & 	 $3.317( -0.6\%)$ & 	 $ 3.337$ \\
\textbf{MgO}    &  $4.99( -4.1\%)$ & 	 $5.17(-0.6\%)$ & 	 $4.878( -6.2\%)$  & 	 $4.834( -7.1\%)$  & 	 $5.110( -1.8\%)$ & 	 $5.057( -2.8\%)$ & 	 $ 5.203$ \\
\textbf{LiF}    &  $4.33( -2.8\%)$ & 	 $4.38(-1.7\%)$ & 	 $4.314( -3.2\%)$  & 	 $4.266( -4.3\%)$  & 	 $4.405( -1.2\%)$ & 	 $4.399( -1.3\%)$ & 	 $ 4.457$ \\
\textbf{NaCl}   &  $3.10( -7.1\%)$ & 	 $3.24(-2.9\%)$ & 	 $3.072( -7.9\%)$  & 	 $3.094( -7.3\%)$  & 	 $3.225( -3.4\%)$ & 	 $3.258( -2.4\%)$ & 	 $ 3.337$ \\
\hline
\textbf{MD} 		 & $-0.175$ 		 & $0.144$ 		 & $-0.219$ 		 & $-0.44$ 		 & $0.123$ 		 & $-0.147$ 		 &  \\
\textbf{MAD} 		 & $0.218$ 		 & $0.206$ 		 & $0.271$ 		 & $0.449$ 		 & $0.195$ 		 & $0.165$ 		 &  \\
\textbf{RMSD} 		 & $0.301$ 		 & $0.275$ 		 & $0.34$ 		 & $0.609$ 		 & $0.258$ 		 & $0.228$ 		 &  \\
\hline
$^a$ Ref. \onlinecite{Tran19}
\end{tabular}
\end{ruledtabular}
\end{table}

\begin{table}[h]
\caption{\label{tab:bmod} Performance assessment on bulk modulus for a set of simple metals, semiconductors and ionic compounds. The table compares 
results in GPa (computed using ONCV-sg15 pseudopotentials at $160$ Ry, fitted for a 4th order polynomial \cite{ZiSc03}) with back-corrected experimental values. The percentage deviation from the experimental values are given in parenthesis.}
\begin{ruledtabular}
\begin{tabular}{llllllll}
System        	& ${\rm PBE}_{\rm ONCV}$ 	&  ${\rm HSE}_{\rm ONCV}$  &  ${\rm CX}_{\rm ONCV}$ & ${\rm AHCX}_{\rm ONCV}$ & ${\rm Exp.}^{a}$ \\
\hline 
\textbf{Cu} 	& 	 $138.0( -4.4\%)$  & 	 $127.7(-11.5\%)$  & 	 $163.3( 13.2\%)$ & 	 $148.3(  2.8\%)$ & 	 $144.3$ \\
\textbf{Ag} 	& 	 $ 87.5(-17.2\%)$  & 	 $ 84.8(-19.8\%)$  & 	 $115.3(  9.1\%)$ & 	 $104.8( -0.9\%)$ & 	 $105.7$ \\
\textbf{Au} 	& 	 $137.4(-24.5\%)$  & 	 $148.5(-18.4\%)$  & 	 $170.5( -6.3\%)$ & 	 $167.8( -7.8\%)$ & 	 $182.0$ \\
\textbf{Pt} 	& 	 $248.5(-13.0\%)$  & 	 $281.1( -1.5\%)$  & 	 $284.0( -0.5\%)$ & 	 $297.7(  4.3\%)$ & 	 $285.5$ \\
\textbf{Rh} 	& 	 $257.6( -7.0\%)$  & 	 $297.1(  7.2\%)$  & 	 $295.8(  6.7\%)$ & 	 $312.7( 12.8\%)$ & 	 $277.1$ \\
\textbf{Al}     & 	 $ 79.5( 10.1\%)$  & 	 $ 79.4( 10.0\%)$  & 	 $ 78.2(  8.3\%)$ & 	 $ 77.3( 7.0\%)$ & 	 $ 72.2$ \\
\textbf{Si}     & 	 $ 87.6(-13.5\%)$  & 	 $ 96.4( -4.8\%)$  & 	 $ 90.1(-11.1\%)$ & 	 $ 96.5( -4.7\%)$ & 	 $101.3$ \\
\textbf{C}      & 	 $432.5( -4.9\%)$  & 	 $470.7(  3.5\%)$  & 	 $439.8( -3.3\%)$ & 	 $466.2(  2.5\%)$ & 	 $454.7$ \\
\textbf{SiC}    & 	 $211.2( -7.8\%)$  & 	 $230.1(  0.4\%)$  & 	 $215.0( -6.2\%)$ & 	 $228.5( -0.3\%)$ & 	 $229.1$ \\
\textbf{GaAs}   & 	 $ 60.3(-21.4\%)$  & 	 $ 73.7( -3.9\%)$  & 	 $ 64.8(-15.5\%)$ & 	 $ 73.8( -3.8\%)$ & 	 $ 76.7$ \\
\textbf{MgO}    & 	 $151.7(-10.7\%)$  & 	 $169.0( -0.5\%)$  & 	 $153.3( -9.7\%)$ & 	 $168.2( -0.9\%)$ & 	 $169.8$ \\
\textbf{LiF}    & 	 $ 67.4(-11.7\%)$  & 	 $ 73.8( -3.3\%)$  & 	 $ 68.3(-10.5\%)$ & 	 $ 74.5( -2.4\%)$ & 	 $ 76.3$ \\
\textbf{NaCl}   & 	 $ 23.7(-14.1\%)$  & 	 $ 24.2(-12.3\%)$  & 	 $ 24.9( -9.8\%)$ & 	 $ 26.1( -5.4\%)$ & 	 $ 27.6$ \\
\hline
\textbf{MD} 	& $-16.9$ 	     & $-3.6$ 		 & $-3.0$ 		 & $3.0$ 		 &  \\
\textbf{MAD} 	& $18.0$ 		 & $10.2$ 		 & $11.2$ 		 & $7.4$ 		 &  \\
\textbf{RMSD} 	& $21.2$ 		 & $14.1$ 		 & $12.4$ 	     & $11.8$ 	     &  \\
\hline
$^a$ Ref. \onlinecite{Tran19}
\end{tabular}
\end{ruledtabular}
\end{table}

\clearpage
%

\end{document}